\documentclass{ar-1col}
 \pdfoutput=1
\usepackage[numbers]{natbib}
\usepackage{url}
\jname{Xxxx. Xxx. Xxx. Xxx.}
\jvol{AA}
\jyear{YYYY}
\doi{10.1146/((please add article doi))}

\begin{document}

\markboth{W.~Li and G.~Wang}{Chiral Magnetic Effects in Nuclear Collisions}

\title{Chiral Magnetic Effects in Nuclear Collisions}

\author{W.~Li$^1$ and G.~Wang$^2$
\affil{$^1$Department of Physics and Astronomy, Rice University, Houston (TX), USA, 77005; email: wl33@rice.edu}
\affil{$^2$Department of Physics and Astronomy, University of California, Los Angeles (CA), USA, 90095; email: gwang@physics.ucla.edu}}

\begin{abstract}
 The interplay of quantum anomalies with strong magnetic field and vorticity in chiral systems could lead to novel transport phenomena, such as the chiral magnetic effect (CME), the chiral magnetic wave (CMW) and the chiral vortical effect (CVE).
 In high-energy nuclear collisions, these chiral effects may survive the expansion of a quark-gluon plasma fireball and be detected in experiments. The experimental searches for the CME, the CMW and the CVE, have aroused extensive interest over the past couple of decades. The main goal of this article is to review latest experimental progress in search for these novel chiral transport phenomena at Relativistic Heavy Ion Collider at BNL and the Large Hadron Collider at CERN. Future programs to help reduce uncertainties and facilitate the interpretation of the data are also discussed.
\end{abstract}

\begin{keywords}
chiral magnetic effect, chiral magnetic wave, chiral vortical effect, heavy-ion collisions, quark-gluon plasma
\end{keywords}
\maketitle

\tableofcontents
\section{Introduction}

An object or system is {\it chiral} if it is not invariant under mirror imaging.
A chiral system carrying an imbalance of right- and left-handed particles can
be characterized by an axial chemical potential ($\mu_5$). In a system of charged chiral fermions with a finite $\mu_5$ value, an electric current can 
be induced when a strong external magnetic field ($\overrightarrow{B}$) is applied,
\begin{equation}
\overrightarrow{J_e} \propto \mu_5\overrightarrow{B}.
\label{eq:CME}
\end{equation}
This phenomenon is known as the chiral magnetic
effect (CME)~\cite{Kharzeev_PLB2006,Kharzeev_NPA2008}. 
The CME has been predicted and observed in condensed matter systems
using Dirac and Weyl semimetals with emergent chiral quasiparticles (e.g., ZrTe$_{5}$~\cite{ZrTe5}, Na$_3$Bi~\cite{Na3Bi}, TaAs~\cite{TaAs} and TaP~\cite{TaP}). In this article, we review the progress of decades-long efforts
in searching for the CME in high-energy nuclear collisions.

In ultra-relativistic heavy-ion (e.g., lead or gold) collisions, a new phase of hot and 
dense nuclear matter is created with a temperature above several trillion
kelvins, consisting of deconfined quarks and
gluons, dubbed the quark-gluon plasma (QGP)~\cite{STAR_white,PHENIX_white,PHOBOS_white,BRAHMS_white}. The chiral symmetry restoration realized in a QGP renders nearly massless or chiral quarks.
It has been suggested that
local chiral domains with finite $\mu_5$ values may be formed in the initial stage of a QGP, via topological charge fluctuations (related to chiral anomaly) in the vacuum of quantum chromodynamics (QCD)~\cite{Kharzeev_PLB2006,Kharzeev_NPA2008,Kharzeev_NPA2007,Kharzeev_PLB2002,Yin_PRL2015,Kharzeev_PRL2010}. Within each domain, there is an imbalance of right- and left-handed chiral quarks (although the global chiral imbalance vanishes after averaging over infinite number of domains). In non-central heavy-ion collisions, extremely strong magnetic fields ($B \sim 10^{15}$~T) can be formed in the QGP, mostly by energetic spectator
protons~\cite{Kharzeev_PLB2006,Kharzeev_NPA2007}. Therefore, the two preconditions (finite $\mu_5$ and $\overrightarrow{B}$ field) for the CME may be realized in heavy-ion collisions, leading to an observable effect such as an electric current along the $\overrightarrow{B}$. The direction of 
$\overrightarrow{B}$ field is approximately perpendicular to the reaction plane (${\rm \Psi_{RP}}$) that contains the impact
parameter and the beam momenta of a collision. As a result, the CME in nuclear collisions will manifest 
an electric charge transport phenomenon across the reaction plane.
The experimental observation of the CME in high-energy nuclear collisions
will have far-reaching impacts in many frontiers of fundamental physics: 
the evolution of the strongest magnetic field ever created, the topological
phases of QCD, and the chiral symmetry restoration of
strong interactions. This prompted a community-wide effort to search for the CME at the Relativistic Heavy Ion Collider (RHIC) in the Brookhaven National Lab (BNL) and
the Large Hadron Collider (LHC) in CERN over the past two decades.

Motivated by the CME search and other modes of collective motions of a QGP, the azimuthal distribution of particles of given transverse momentum ($p_T$) and pseudorapidity ($\eta$) in an event is typically decomposed by a Fourier series:
\begin{equation}
    \frac{dN_{\alpha}}{d\phi} \propto 1 + 2v_{1,\alpha}\cos(\Delta\phi) + 2v_{2,\alpha}\cos(2\Delta\phi) + ... + 2a_{1,\alpha}\sin(\Delta\phi) + ...,
\label{equ:Fourier_expansion}
\end{equation}
\noindent where $\phi$ is the azimuthal angle of a particle, and $\Delta\phi = \phi - {\rm \Psi_{RP}}$.
The subscript $\alpha$ ($+$ or $-$) denotes the charge sign of a particle.
Conventionally, coefficients $v_1$, $v_2$ and $v_3$ of parity-even terms are called ``directed flow", ``elliptic flow", and ``triangular flow", respectively. They 
reflect the hydrodynamics response of the QGP medium to the initial collision geometry and to its fluctuations, respectively~\cite{HYDRO_review}.
The coefficient $a_1$ (with $a_{1,-} = -a_{1,+}$) of the parity-odd term quantifies the electric charge separation with respect to the reaction plane, e.g., due to the CME.

Another transport phenomenon complementary to the CME is the chiral separation effect (CSE)~\cite{CSE1,CSE2},
in which a current of chiral charges along the $\overrightarrow{B}$ field 
is induced by a finite chemical potential of vector charges (e.g., electric charges):
\begin{equation}
    \overrightarrow{J_5} \propto \mu_{\rm v}\overrightarrow{B}.
\end{equation}
\noindent The interplay of the CME and CSE forms a collective 
excitation, the chiral magnetic wave (CMW), which is a long wavelength hydrodynamic mode of chiral charge densities~\cite{CMW,CMW2}.
The CMW can manifest itself as an electric quadrupole moment of the
collision system, where the ``poles" (``equator") of the produced fireball acquire additional positive (negative) charges~\cite{CMW}.
This effect can be explored in the measurements of charge-dependent elliptic flow ($v_2$). Observation of the CMW does not necessarily depend on the observation of the CME, as the latter requires an initial $\mu_5$ from the QCD chiral anomaly (which may be small), while the former only needs a local net electric charge density.

Finally, in analogy to the $\overrightarrow{B}$-field induced
anomalous chiral transport effects, similar phenomena can also take 
place when a chiral system carries a global angular momentum under rotation. 
The fluid rotation can be quantified by vorticity, 
$\overrightarrow{\omega}=\overrightarrow{\bigtriangledown}\times\overrightarrow{v}$, where
$\overrightarrow{v}$ is the flow velocity field. Given a large vorticity,
the chiral vortical effect (CVE)~\cite{CVE} can induce a vector current:
\begin{equation}
    \overrightarrow{J_{\rm v}}\propto \mu_5\mu_{\rm v}\overrightarrow{\omega}.
\end{equation}
While the CME is driven by $\overrightarrow{B}$, the CVE is driven by $\mu_{\rm v}\overrightarrow{\omega}$ of the QGP.
Here the subscript ``v" denotes ``vector", which can be, e.g., ``$B$" (baryonic charge) or ``$e$" (electric charge). The $\mu_B$ is not
affected by the presence of $\overrightarrow{B}$ field, 
making it a better tool to search for the CVE through the baryonic-charge separation in heavy-ion collisions.
In the case of the CVE search, the subscript $\alpha$ in Eq.~\ref{equ:Fourier_expansion} would represent baryon or anti-baryon numbers.

There are more chiral magnetic/vortical effects proposed, such as the chiral electric separation effect (CESE)~\cite{CESE1,CESE2} and
the chiral vortical wave (CVW)~\cite{CVW} (see Ref~\cite{Jinfeng} for a review on these effects).

This article focuses on the key experimental results over the
past couple of decades in search for the chiral magnetic/vortical effects
in high-energy nuclear collisions: the probe of the initial magnetic field and vorticity in Section~\ref{sec:Drive}, experimental searches for the electric-(baryonic-)charge separation in Section~\ref{sec:CME} (Section~\ref{sec:CVE}), and the electric quadrupole moment in Section~\ref{sec:CMW}. An outlook for future developments is discussed in Section~\ref{sec:Outlook}.

\section{Magnetic field and vorticity in nuclear collisions}
\label{sec:Drive}

Intuitively, one may regard the magnetic field (vorticity) as the driving force of the CME (CVE),
the chirality imbalance as the initial condition, and the electric-(baryonic-)charge separation as the manifestation. The existence and magnitude of the magnetic field (vorticity) could be independently constrained by other experimental observable.

The initial magnetic field is roughly estimated to be, 
\begin{equation}
eB \sim \gamma \alpha_{\rm EM} Z / b^2,    
\end{equation}
where $\alpha_{\rm EM} \simeq 1/137$, $b$ is the impact parameter, and $\gamma$ is the Lorentz factor. The large charge number, $Z$, and small impact parameter, $b$, lead to an extremely strong magnetic field.
A typical noncentral Au+Au collision at $\sqrt{s_{\rm NN}}= 200$ GeV produces an initial $eB \sim 1/(1{\rm fm}^2) \sim m_{\pi}^2$ (or $10^{14}$ T).
Many model calculations have attempted to quantify the electromagnetic field in detail on the event-by-event basis,
in terms of its spatial distribution, the fluctuation of its orientation as well as the dependence on colliding nuclei, 
centrality and beam energy (see Refs.~\cite{mag1,mag2,mag3} for examples).
A major uncertainty in calculations of the magnetic field $\overrightarrow{B}$ is 
its lifetime during the evolution of the QGP created in heavy-ion collisions (see Refs.~\cite{mag4,mag5,mag6,mag7} for examples).
The time dependence of $\overrightarrow{B}$ after the initial impact crucially depends on  whether/when/how a conducting medium may be formed. In the vacuum, the magnetic field will
quickly decay as two ions pass by, whereas if the QGP medium carries an electric conductivity, a longer duration may be sustained.

The electric conductivity~\cite{conductivity} and the time evolution of quark densities~\cite{PHSD} 
can be studied via charge-dependent directed flow of final-state hadrons in asymmetric A+A collisions, such as Cu+Au.
The difference in the number of protons between Au and Cu creates a strong electric field in the initial stage of the collision,
pointing from Au to Cu. The lifetime
of the electric field might be very short (e.g. $t\sim0.25$ fm/$c$ depending on conductivity 
from Refs.~\cite{conductivity,PHSD}). If quarks and antiquarks are produced sufficiently early in the collision, they would experience a Coulomb force, and the degeneracy in $v_1$ is lifted between positively and negatively charged particles~\cite{mag2,conductivity}:
\begin{equation}
v_{1}^{\pm} = v_{1} \pm d_{E}\langle\cos({\rm \Psi_{RP}} - {\rm \Psi}_{E})\rangle,    
\end{equation}
where ${\rm \Psi}_E$ denotes the azimuthal angle of the electric field,
and the coefficient $d_E$ characterizes the strength of dipole deformation induced by the electric
field, proportional to the electric conductivity of the medium.
Here $v_1$ represents the rapidity-even component of directed flow
that dominates over the rapidity-odd one in asymmetric collisions. In symmetric collisions, $v_1$ often denotes the latter.

The STAR collaboration has measured  charge-dependent $v_1^{\rm even}$ and the difference $\Delta v_1^{\rm even}$
as functions of $p_T$ in both Cu+Au and Au+Au collisions at 200 GeV~\cite{v1@CuAu,v1@AuAu}. 
For $p_T < 2$ GeV/$c$, the $\Delta v_1^{\rm even}$ seems to increase with $p_T$ in Cu+Au collisions, while it is consistent with zero in Au+Au collisions.
The parton-hadron-string-dynamics (PHSD) model~\cite{PHSD} 
is a dynamical transport approach in the partonic and hadronic phases, and has calculated the $\Delta v_1^{\rm even}$ 
driven by the initial electric field.  
The model assumes that all electric charges are affected by the electric field, which results in a large separation
of $v_1^{\rm even}$ between positive and negative particles. 
After scaling down the calculated $\Delta v_1^{\rm even}$ by a factor of 10, the model describes rather well the $p_T$ dependence of experimental data for $p_T < 2$ GeV/$c$.
This qualitative observation of the strong initial electric field in asymmetric collisions
provides an indirect evidence for the strong initial magnetic field in heavy-ion collisions,
which shares the same conducting medium with the electric field, and hence
could also leave an imprint on the final-stage particles.

Heavy-flavor quarks, such as {\it charm (c)} and {\it beauty (b)}, are produced much earlier than light-flavor quarks in a collision because of their large masses. As a result, heavy quarks have a better chance to witness the strong electric and magnetic fields~\cite{EMField_D0,EMField_vn}. In symmetric A+A collisions, it has been predicted that a rapidity-dependent splitting of $D^{0}$ and $\bar{D}^{0}$ meson 
$v_1$ and $v_3$ (from the magnetic field), and $v_2$ (from the electric field) will be generated. The same phenomenon is expected to occur for light hadrons but at a much reduced magnitude because of the much later production time. Experimental efforts are on
ongoing to accumulate high-precision data sets to explore these effects. 

Vorticity  results from the interplay of global rotation and shear viscosity of the QGP in heavy-ion collisions.
In a noncentral collision, the majority of the global angular momentum, $\overrightarrow{L}$, 
is carried away by spectator nucleons.
However, about $10-20\%$ of $\overrightarrow{L}$ could remain in the QGP and
be approximately conserved over time~\cite{vor1,vor2}.
This implies that the CVE can be developed over a relatively long duration.  The angular momentum is largely aligned with the magnetic-field direction, both perpendicular to the reaction plane,
so the CME and CVE are very much alike in terms of their experimental observables.
Attempts to compute local vorticity $\overrightarrow{\omega}$ and its space-time
distribution have also been made extensively~\cite{vor1,vor2,vor3,vor4,vor5,vor6}.

Experimentally, the global polarization of hyperons such as the $\Lambda$ baryon has been used to probe both the QGP vorticity and the magnetic field.
The local vortical effects can generate a positive spin polarization for both
$\Lambda$ and $\bar{\Lambda}$, whereas the coupling of the hadronic magnetic dipole moment
to the magnetic field will produce a negative (positive) contribution for $\Lambda$ ($\bar{\Lambda}$).
Therefore, observing a splitting between $\Lambda$ and $\bar{\Lambda}$ polarization will be a direct evidence for the magnetic field. Note that $\Lambda$ is typically produced in a later stage of the collision, so its sensitivity to the initial magnetic field could be limited by the lifetime of the magnetic field. The first observation of global
$\Lambda$ and $\bar{\Lambda}$ polarization in heavy-ion
collisions has been reported by the STAR
collaboration~\cite{Isaac}. At $\sqrt{s_{\rm NN}}<100$ GeV,
the signal is on the order of a few percent, and displays some hint of a weak beam-energy dependence.
Current statistical precision of data is insufficient to study polarization for $\Lambda$ and $\bar{\Lambda}$ separately.
This question will be addressed in
the second beam energy scan (BES-II) program at 
RHIC~\cite{BESII} to search for evidence of the magnetic field. 

Further searches for  the initial magnetic field (vorticity) have been proposed through photon (vector meson)
polarization measurements~\cite{polarization}. The initial magnetic helicity ($\overrightarrow{E}\cdot\overrightarrow{B}$) of the collision system can be quite large, 
with opposite signs in the upper and lower hemispheres. Because of the chiral anomaly, helicity can be
transferred back and forth between the magnetic flux and fermions, so that the magnetic
helicity could last long enough to yield photons with opposite circular polarizations in the hemispheres above and below
the reaction plane~\cite{Yin,Ipp,Yee,Mamo}. 
The initial global quark polarization
could effectively lead to a polarization of photons~\cite{Ipp},
and hence cause an 
asymmetry in photon polarization~\cite{XNWang}.
This local imbalance of photon circular polarization could be investigated via the polarization
preference with respect to the reaction plane for photons that convert into $e^+e^-$ pairs~\cite{polarization}.
Similarly, vector mesons that decay into two daughters can also have their polarization preferences measured
with the scheme outlined in Ref.~\cite{polarization}, and the helicity separation in this case
originates from vorticity~\cite{XNWang,Baznat,XGHuang}.

Another proposal aiming at measuring the imprint left by the initial
magnetic field focuses on pairs of oppositely charged particles in the whole evolution of heavy-ion collisions~\cite{YuGang}. The pertinent mechanism is the distortion of the relative angle between
positively- and negatively-charged particles inside a pair.
Two observables are adopted to detect this effect: one based on the same framework as that measuring global $\Lambda$  polarization, and the other based on a slightly modified balance function.  The knowledge documented in Ref~\cite{YuGang} will facilitate the experimental efforts to quantify the strong magnetic field in high-energy
nuclear collisions.
\section{CME searches in nuclear collisions}
\label{sec:CME}
In this section, methodologies employed to search
for the CME (as well as other anomalous chiral effects) in nuclear collisions are reviewed, followed by experimental results at RHIC~\cite{STAR1,STAR2,STAR3,STAR4,STAR5,STAR6,UU1,Prithwish,PHENIX1,PHENIX2} and the LHC~\cite{ALICE,CMS1,CMS2}. 
These include data in large A+A collision systems like
Au+Au, U+U and Pb+Pb, and also small systems such as p+Au, d+Au and p+Pb. In particular, direct comparison of large- and small-system data in recently years are proven to provide a powerful tool in better understanding background contributions. Approaches developed for quantifying
the background contributions and extracting true
CME signals are also discussed. In the outlook, prospects of future programs at RHIC for isobaric collisions and the Beam Energy Scan (BES-II), and at the high-luminosity LHC are reviewed.

\subsection{Methodologies}

Measurements of the CME-induced charge separation across the
reaction plane are primarily explored by the so-called {\it 
three-point $\gamma$ correlator}, first proposed in 
Ref.~\cite{Voloshin:2004vk}.

It is tempting to directly measure the event-averaged $a_{1,\pm}$
coefficient from the single-particle azimuthal 
distribution in Eq.~\ref{equ:Fourier_expansion}. 
However, since the sign of the $\mu_5$ value fluctuates between positive and negative on an event-by-event basis with equal probability (global parity for QCD should be conserved), the event-averaged $a_{1,\pm}$ values are zero by construction. 
The $\gamma$ correlator (later often referred to as $\gamma_{112}$) is designed to observe the fluctuations of
charge separations or $a_{1,\pm}$ coefficients
with respect to the reaction plane~\cite{Voloshin:2004vk},
\begin{eqnarray}
\gamma &\equiv& \langle \langle \cos(\phi_\alpha + \phi_\beta -2{\rm \Psi_{RP}}) \rangle\rangle \nonumber \\
&=& \langle\langle\cos(\Delta\phi_{\alpha})\cos(\Delta\phi_{\beta}) -
\sin(\Delta\phi_{\alpha})\sin(\Delta\phi_{\beta})\rangle\rangle \nonumber \\
&=& (\langle v_{1,\alpha}v_{1,\beta}\rangle + B_{\rm IN}) -(\langle a_{1,\alpha}a_{1,\beta}\rangle + B_{\rm OUT}), \label{eq:ThreePoint},
\end{eqnarray}
\noindent where the averaging is done over all combinations of particle $\alpha$ and $\beta$ in an event and over all events.
The expansion of the $\gamma$ correlator
reveals the difference between {\it in-plane} and {\it out-of-plane} projections of azimuthal correlations.
The third term of Eq.~\ref{eq:ThreePoint}, $\langle a_{1,\alpha}a_{1,\beta}\rangle$,
represents a measurement of the variance (or fluctuations) of $a_{1,\pm}$ coefficients, which is the main target for the CME search. There are other terms that are presumably unrelated to the CME. 
The first term, $\langle v_{1,\alpha}v_{1,\beta}\rangle$, is related to the
directed flow that is expected to be charge independent and unrelated to 
the magnetic field in symmetric A+A systems. The $B_{\rm IN}$ and $B_{\rm OUT}$ terms represent other
possible background correlations (as will be discussed in detail later) in and out-of the reaction plane, respectively.
By taking a difference between opposite-sign and same-sign $\gamma$ correlators,

\begin{equation}
\Delta \gamma \equiv \gamma^{\rm OS} - \gamma^{\rm SS}, 
\end{equation}

\noindent the $\langle v_{1,\alpha}v_{1,\beta}\rangle$ terms cancel out, as well as a large portion of the background terms ($B_{\rm IN}$ and $B_{\rm OUT}$) that are reaction-plane independent.
There may still be a residual reaction-plane dependent background in ($B_{\rm IN}-B_{\rm OUT}$), at a level proportional to the magnitude of elliptic flow coefficient ($v_2$).  This is the major unknown source of backgrounds in $\Delta \gamma$ measurements. 
In practice, the reaction plane is approximated with the ``event plane" ($\rm \Psi_{EP}$) reconstructed with measured particles,
and then the measurement is corrected for the finite event plane resolution.
The main advantages of the $\gamma$ correlator lie in its direct connection to the $a_1$ coefficient and a relative straight forward procedure for correcting the event plane resolution.

Several alternative methods to the $\gamma$ correlator were also proposed, with the goal of 
providing complementary sensitivity to the CME signal and backgrounds. These include the modulate sign correlator (MSC)~\cite{STAR3}, the charge multiplicity asymmetry correlator (CMAC)~\cite{STAR4}, the multi-particle charge-sensitive correlator ($R_{{\rm \Psi}_m}(\Delta S)$)~\cite{PHENIX1,PHENIX2,Roy}, and the signed balance functions~\cite{Aihong_BF}. It is not a surprise that these methods provide
largely overlapping information with the $\gamma$ correlator, as they all 
make use of the same inputs of particle azimuthal correlations. These alternative methods
are mostly in the process of being applied to experimental data.
Therefore, we will focus on reviewing experimental results of the $\gamma$ correlator 
and its derivatives in the following discussion.

\subsection{Results in large A+A systems: evidence for the CME}

\begin{figure}[th]
\includegraphics[width=0.5\textwidth]{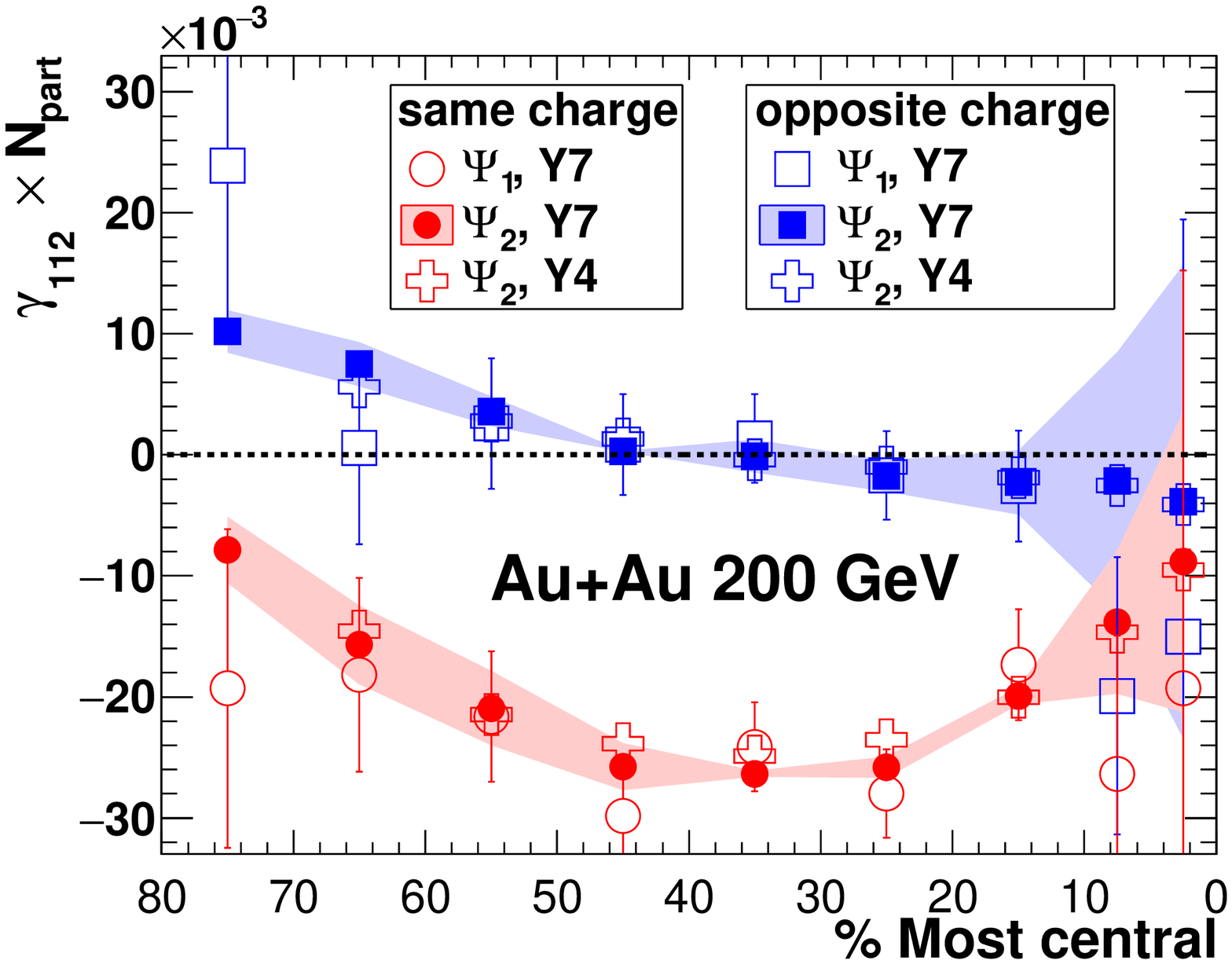}
\caption{Charge-dependent $\gamma_{112}\times N_{\rm part}$ measured with the 1st-order spectator plane ($\Psi_1$) and the 2nd-order
participant plane ($\Psi_2$) versus centrality for Au+Au collisions
at 200 GeV~\cite{STAR1,STAR2,STAR3}. Y4 and Y7 represent STAR results from the RHIC runs 2004 and 2007, respectively.}
\label{fig1}
\end{figure}

Measurements of the three-point $\gamma_{112}$ correlator 
have been performed extensively in Au+Au collisions over a wide range of RHIC energies and in Pb+Pb collisions at top LHC energies.
Figure~\ref{fig1} shows the result of opposite-sign and same-sign correlators ($\gamma_{112}\times N_{\rm part}$), respectively, 
in Au+Au collisions at 200 GeV, measured by the STAR collaboration~\cite{STAR1,STAR2,STAR3}. Here, the number of participating nucleons, $N_{\rm part}$, is used as a multiplicative factor to 
compensate for the expected dilution of signals with increasing number of
domains (with random signs of $\mu_5$) toward more central Au+Au collisions. In this way, results become less centrality dependent.

An apparent charge dependence of the three-point correlator is 
observed, where values of $\gamma_{112}^{\rm OS}$ ($\sim$ 0) are generally higher than those of $\gamma_{112}^{\rm SS}$ ($<$ 0). This observation was regarded as the evidence for the CME, resulting in 
collective charge separations with respect to the reaction plane. The observation is robust against different approaches of reconstructing the event plane.
If the CME-induced charge separation is the only physical 
origin of the $\gamma_{112}$ correlations, $\gamma_{112}^{\rm OS}$ and $\gamma_{112}^{\rm SS}$ should have the same magnitude but opposite signs, symmetric around zero. This is obviously not the case in data, indicating that charge-independent backgrounds   must be present (e.g., momentum conservation and collective flow effects).

\begin{figure}[th]
\includegraphics[width=\textwidth]{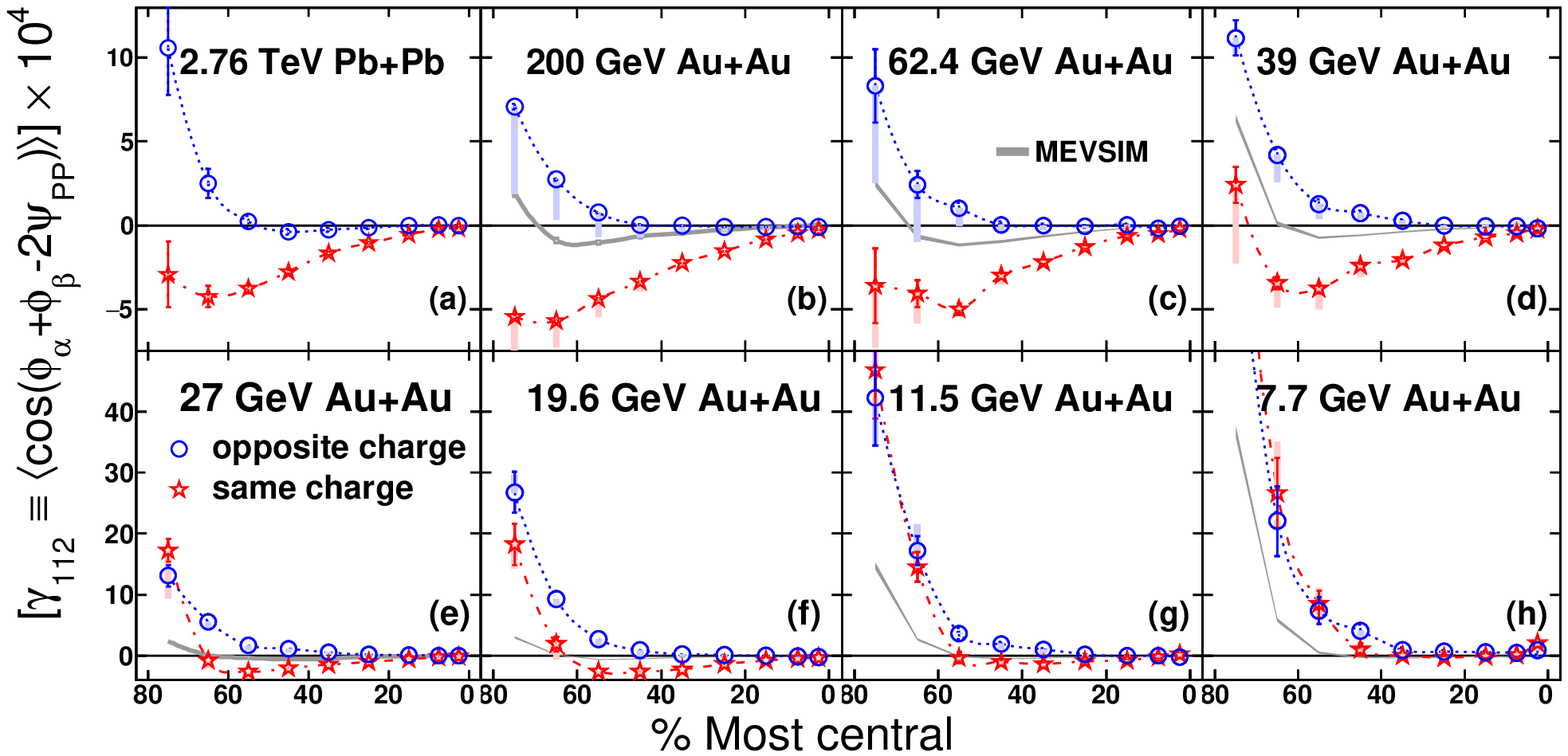}
\caption{Charge-dependent $\gamma_{112}$ measured with the the 2nd-order
participant plane ($\Psi_{\rm PP}$) versus centrality for Pb+Pb collisions at 2.76 TeV~\cite{ALICE}, and for Au+Au collisions
at 200, 62.4, 39, 27, 19.6, 11.5 and 7.7 GeV~\cite{STAR5}.}
\label{fig2}
\end{figure}

Following the first observation at 200 GeV, the charge-dependent $\gamma_{112}$ correlator has been measured over a wide range of
collision energies at RHIC and the LHC, shown in Fig.~\ref{fig2}, as a function of centrality for Pb+Pb collisions at 2.76 TeV~\cite{ALICE}, and for Au+Au collisions
at 200, 62.4, 39, 27, 19.6, 11.5 and 7.7 GeV~\cite{STAR5}.
The charge-independent backgrounds tend to be 
more prominent in lower beam energies, where the multiplicity is lower.
The MEVSIM model calculation~\cite{MEVSIM} with momentum conservation effects but no CME can qualitatively
capture this feature of experimental data.
The difference between $\gamma_{112}^{\rm OS}$ and
$\gamma_{112}^{\rm SS}$ persists up to the\ LHC energies and down to the RHIC BES energies, while there is a hint of diminishing at 7.7 GeV. To focus on the charge-dependent correlation signals and isolate
background effects that are most relevant to CME searches, 
the difference
between $\gamma_{112}^{\rm OS}$ and $\gamma_{112}^{\rm SS}$
is also often studied. 

\subsection{Approaches to disentangle the signal vs. background}

{\it Extraordinary claims require extraordinary evidence.}
While the A+A data have provided evidence in line with the observation of the CME, no definitive conclusion can be drawn yet because of the presence of several charge-dependent background contributions. A class of main charge-dependent backgrounds can be generalized into
the local charge conservation (LCC) or ordering effect in decays of resonance and/or cluster-like intermediate states. 
These resonances and clusters develop a correlation with the reaction plane during the anisotropic hydrodynamic expansion of the QGP, seeded by an asymmetric initial geometry.
Precision theoretical calculations of those background contributions 
are not available, because of their nonperturbative nature. Data-driven approaches and phenomenological models have been 
developed to quantitatively constrain the backgrounds 
and examine if it is necessary to invoke the CME mechanism to explain the experimental data. 
For example, a model calculation incorporating the LCC effect and 
anisotropy flow~\cite{PrattSorren:2011} was able to 
capture the STAR data without introducing any CME signal.

A generic scenario of anisotropic cluster emission (also called ``flowing clusters'') were originally considered in Ref.~\cite{Voloshin:2004vk} to investigate the background terms $B_{\rm IN}$ and $B_{\rm OUT}$ in the $\gamma$ correlator:
\begin{equation}
    \frac{B_{\rm IN}-B_{\rm OUT}}{B_{\rm IN}+B_{\rm OUT}} \approx v_{2,{\rm cl}} \frac{\langle\cos(\phi_\alpha+\phi_\beta-2\phi_{\rm cl})\rangle}{\langle\cos(\phi_\alpha-\phi_\beta\rangle)},
\end{equation}
where $\phi_{\rm cl}$ is the cluster emission azimuthal angle, and $\phi_\alpha$ and $\phi_\beta$ are
the azimuthal angles of two decay products. The
$v_{2,{\rm cl}}$ of clusters contains both flow and so-called nonflow contributions (e.g., short-range correlations within a cluster).
The flowing cluster model can be generalized to a larger portion of or even the full event,
through the mechanisms of transverse momentum conservation (TMC)~\cite{Pratt2010,Flow_CME} 
and/or local charge conservation (LCC)~\cite{PrattSorren:2011}.
Ideally, the two-particle correlator, 
$\delta \equiv \langle \cos(\phi_\alpha -\phi_\beta) \rangle$, should be proportional to $\langle a_{1,\alpha} a_{1,\beta} \rangle$,
but in reality it is strongly dominated by short-range two-particle correlation backgrounds.
For example, the TMC effect leads to the following pertinent correlation terms in $\Delta \delta$ and $\Delta \gamma_{112}$~\cite{Flow_CME}:
\begin{eqnarray}
\Delta \delta^{\rm TMC} &\rightarrow& -\frac{1}{N}
\frac{\langle p_T \rangle^2_{\rm \Omega}}{\langle p_T^2 \rangle_{\rm F}}
\frac{1+({\bar v}_{2,{\rm \Omega}})^2-2{\bar{\bar v}}_{2,{\rm F}}{\bar v}_{2,{\rm \Omega}}} {1-({\bar{\bar v}}_{2,{\rm F}})^2},
\\
\Delta \gamma^{\rm TMC}_{112} &\rightarrow& -\frac{1}{N}
\frac{\langle p_T \rangle^2_{\rm \Omega}}{\langle p_T^2 \rangle_{\rm F}}
\frac{2{\bar v}_{2,{\rm \Omega}}-{\bar{\bar v}}_{2,{\rm F}}-{\bar{\bar v}}_{2,{\rm F}}({\bar v}_{2,{\rm \Omega}})^2} {1-({\bar{\bar v}}_{2,{\rm F}})^2}
\nonumber \\
&\approx& \kappa^{\rm TMC}_{112} \cdot v_{2,{\rm \Omega}} \cdot \Delta \delta^{\rm TMC},
\label{eq:11}
\end{eqnarray}
where $\kappa^{\rm TMC}_{112} = (2{\bar v}_{2,{\rm \Omega}}-{\bar{\bar v}}_{2,{\rm F}})/v_{2,{\rm \Omega}}$,
and ${\bar v}_{2}$ and ${\bar{\bar v}}_{2}$ represent the $p_T$- and $p_T^2$-weighted moments of $v_2$, respectively.
The subscript ``F" denotes an average of all produced particles in the full phase space;
the actual measurements will be only in a fraction of the full space, denoted by ``${\rm \Omega}$".
The background contribution due to the LCC effect has a similar characteristic structure
as the above~\cite{Pratt2010,PrattSorren:2011}. This motivates a normalization of $\Delta \gamma$ by $v_2$ and $\Delta \delta$:
\begin{equation}
    \kappa_{112} \equiv \frac{\Delta \gamma_{112}}{v_2 \cdot \Delta \delta}.
\label{kappa112}
\end{equation}
Only when $\kappa_{112}$ is larger than $\kappa^{\rm TMC}_{112}$, may a CME signal be present. In Au+Au collisions at 200 GeV, the $\kappa^{\rm TMC}_{112}$ values estimated in the model using PHOBOS $v_2$ data~\cite{PHOBOS1,PHOBOS2} are around $1.3$ for all the available centrality intervals, and the $\kappa_{112}$ values from background-only simulations of A Multi-Phase Transport (AMPT) 
model~\cite{ampt1,ampt2,ampt3} show
a seemingly constant of $1.3$ over the $0-80\%$ centrality range~\cite{Subikash}. On the other hand, the STAR data for such collisions typically
bear $\kappa_{112} \ge 2$~\cite{Kong}, implying that one third of the $\Delta \gamma_{112}$ correlation could arise from the CME. However, the AMPT model may not capture all background contributions, and thus cannot be relied on to quantify the CME signal contribution.

\begin{figure}[h]
\includegraphics[width=2.5in]{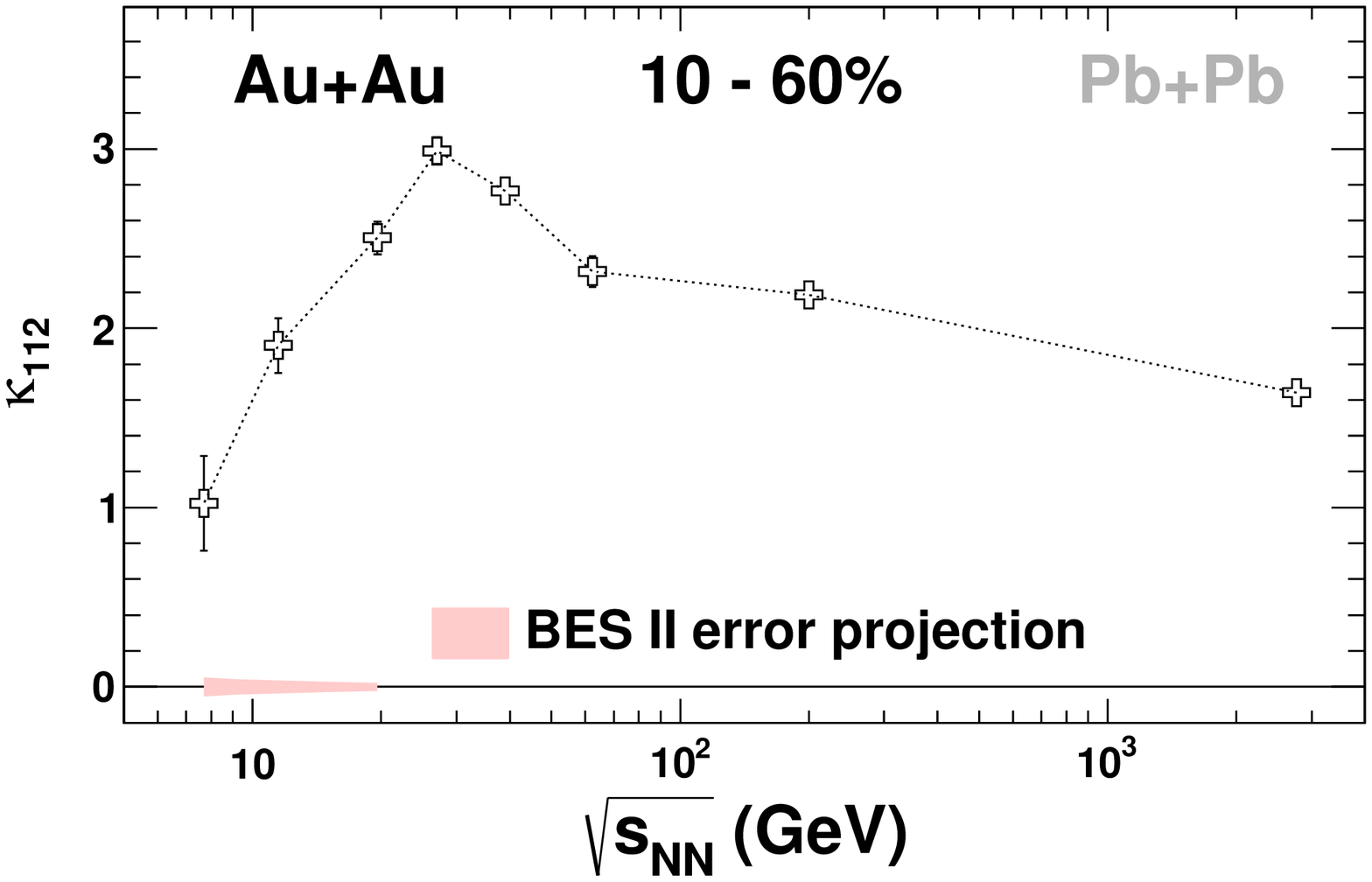}
\caption{$\kappa_{112}$ measured with the participant plane versus beam energy for 10-60\% Pb+Pb collisions at 2.76 TeV~\cite{ALICE}, and for 10-60\% Au+Au collisions
at 200, 62.4, 39, 27, 19.6, 11.5 and 7.7 GeV~\cite{STAR5}. 
The error projection for the RHIC BES II is also presented.}
\label{fig3}
\end{figure}

Figure~\ref{fig3} shows the beam-energy dependence of 
$\kappa_{112}$
for 10-60\% Pb+Pb collisions at 2.76 TeV~\cite{ALICE}, and for 10-60\% Au+Au collisions
at 200, 62.4, 39, 27, 19.6, 11.5 and 7.7 GeV~\cite{STAR5}. The data display a rise-and-fall trend, with a peak around 27 GeV Au+Au collisions
and a drop approaching the background level at 7.7 GeV.
The disappearance of the CME is expected at low collision energies where the partonic interactions are dominated by the hadronic ones,
and quarks are not massless any more. At LHC energies, although the initial magnetic field has a much stronger peak magnitude than at RHIC, it drops more rapidly, possibly vanishing before the formation of the QGP. Without any electric conductivity in the QGP, by the time scale of 0.1 fm/$c$, the remaining magnetic field at LHC is lower than that at RHIC typically by two orders of magnitude. Therefore, a smaller CME coud be anticipated at LHC than at RHIC energies.

Data-driven approaches to constrain the background contributions are discussed below, which have the advantage of being model independent. These are general strategies: (1) to vary the signal while keeping the background fixed, such as using small systems and higher-order $\gamma$ correlators; (2) to vary the background while keeping the signal fixed, such as the
event shape engineering.

\subsubsection{Results in small p(d)+A systems}

\begin{figure}[th]
\includegraphics[width=0.7\textwidth]{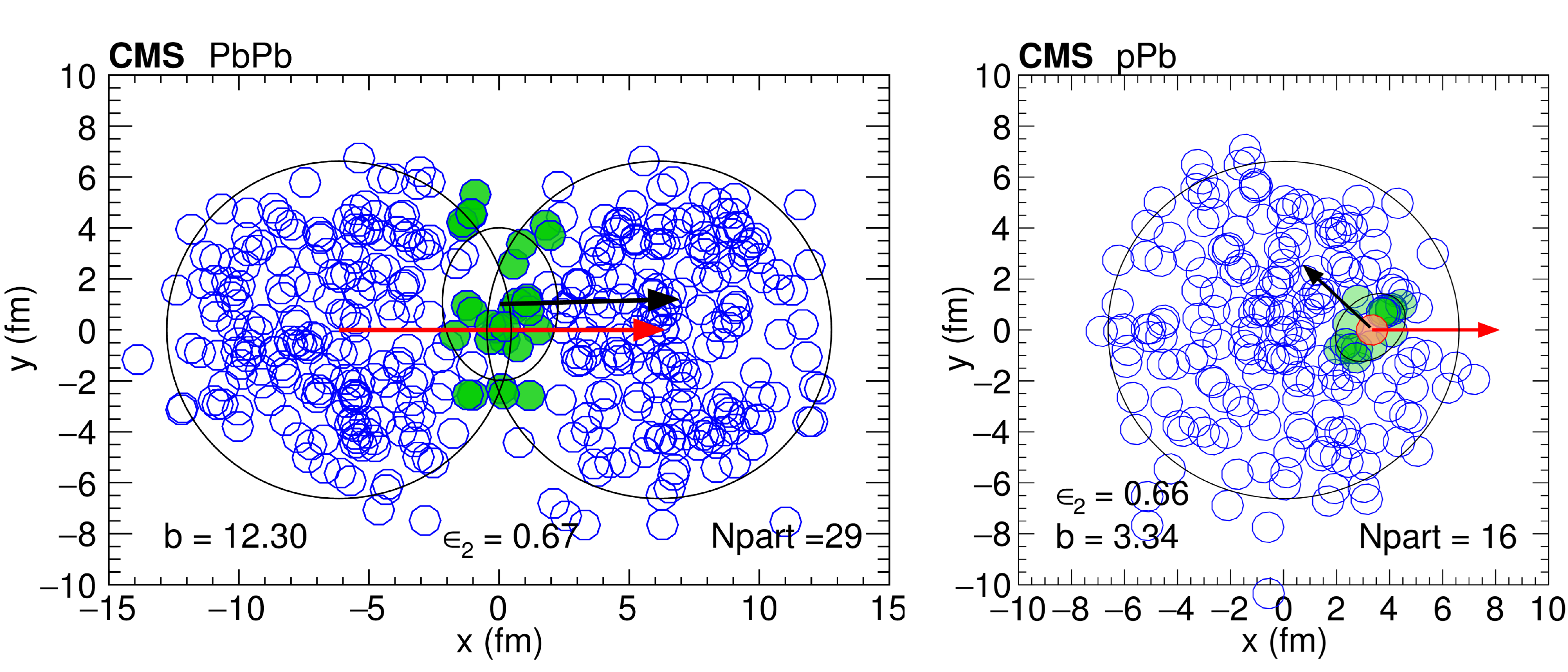}
\caption{Cartoons for demonstrating the correlation (and the decorrelation) between the participant plane (black arrows) and the reaction plane (red arrows) in A+A (left) and p+A (right) collisions.}
\label{fig:cme_demo_small}
\end{figure}

In a non-central A+A collisions, the participant plane of
the lenticular overlap region is, although fluctuating,
generally strongly correlated with the reaction plane,
or perpendicular to the magnetic field, as illustrated in
Fig.~\ref{fig:cme_demo_small} (left) for a Pb+Pb collision.
Conversely, in a p+Au(Pb) collision, the overlapping
geometry is entirely determined by fluctuations, and the
participant plane is essentially uncorrelated with the reaction
plane or the magnetic field direction, as illustrated in
Fig.~\ref{fig:cme_demo_small} (right). Therefore, even if the 
magnetic field may still be comparable in magnitude to A+A systems in such small-system collisions, 
its decoupling from the participant plane will greatly 
suppress the possible CME contribution in $\gamma_{112}$.
Meanwhile, it has been observed in recent years that small systems exhibit similar bulk properties to large A+A systems~\cite{SmallSystem_Li,SmallSystem_Nagle}. All these make the small system an ideal data-driven testing ground for turning off the possible CME signal and understanding the pure-background contributions.

\begin{figure}[th]
\includegraphics[width=\textwidth]{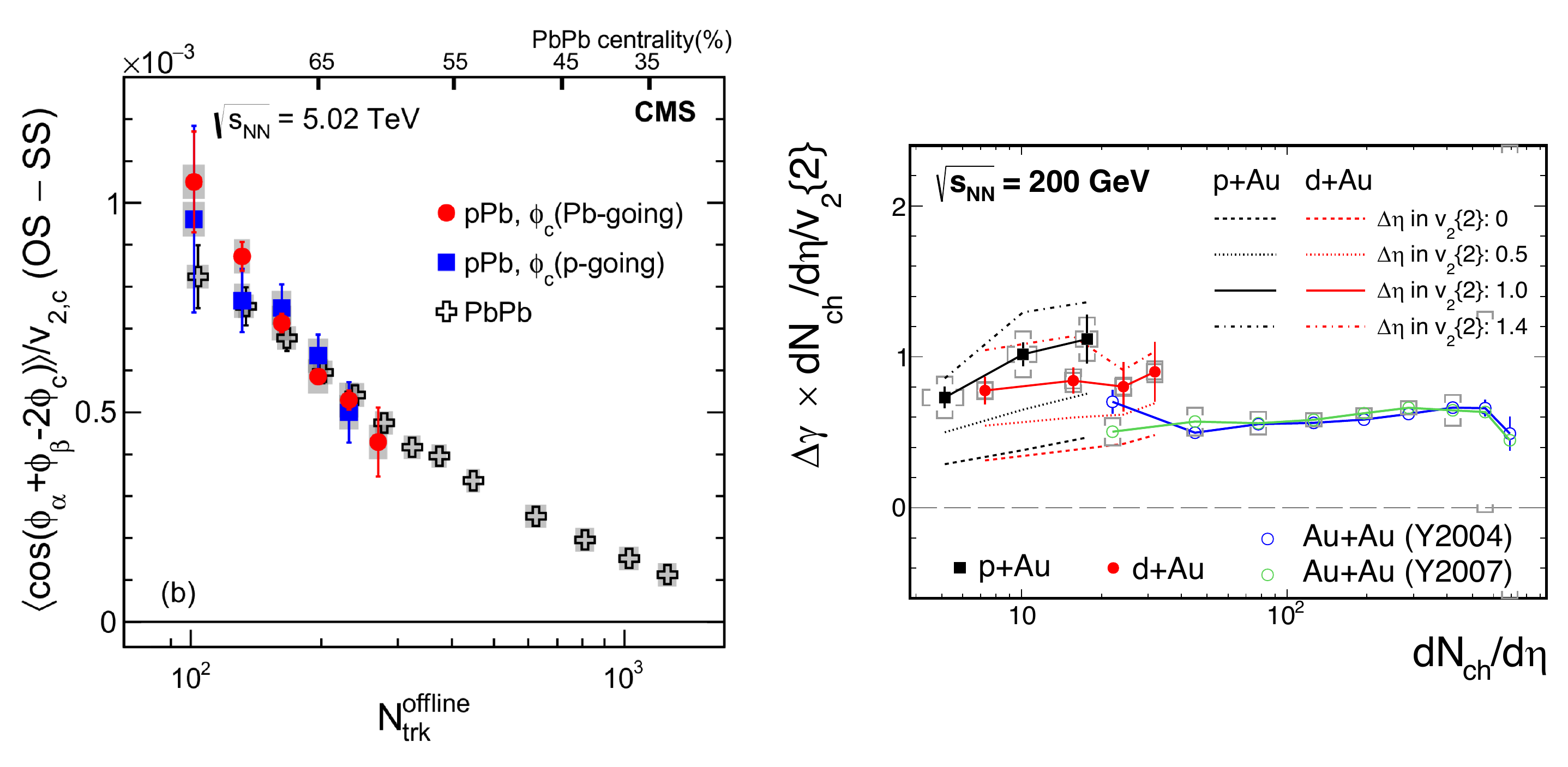}
\caption{The opposite-sign (SS) and same-sign (OS) difference $\Delta\gamma_{112}$ correlator as a function of event multiplicity in p+Pb and Pb+Pb collisions at 5.02 TeV from CMS~\cite{CMS1} (left) and
p+Au, d+Au and Au+Au collisions at 200 GeV from STAR~\cite{STAR6} (right).
}
\label{fig:cms_pPbPbPb}
\end{figure}

The $\Delta \gamma_{112}$ correlator in p+Pb collisions 
has been measured by CMS~\cite{CMS1}, shown in
Figure~\ref{fig:cms_pPbPbPb} (left) as a function of 
multiplicity, together with Pb+Pb data at 5.02 TeV~\cite{CMS1}.
At the same multiplicity, the p+Pb and Pb+Pb data are nearly identical over a wide range of multiplicities. This observation indicates that
the charge separation signal observed in A+A collisions is likely
to be dominated by, if not entirely, background correlations unrelated
to the CME. In this analysis, a large $\eta$ gap of at least 2 units
is required between particle $\alpha$,$\beta$ and the event plane, 
largely eliminating backgrounds directly from short-range correlations. 
Long-range nonflow correlations such
as di-jets are also present and their effects tend to diminish as event
multiplicity increases or from peripheral to most central A+A collisions.
Although small-system data strongly indicate the dominance of backgrounds
up to semi-peripheral A+A events, caution should be taken when
extrapolating to more central collisions, where contributions of
different physics processes may vary.

STAR has also measured the $\Delta \gamma_{112}$ correlator in small systems of p+Au and d+Au collisions at 200 GeV~\cite{STAR6}. The results of $\Delta\gamma_{112}$ scaled by $dN_{\rm ch}/d\eta/v_2$ are shown in Fig.~\ref{fig:cms_pPbPbPb}. An interesting ordering of $\Delta\gamma_{112}^{pAu} > \Delta\gamma_{112}^{dAu} > \Delta\gamma_{112}^{AuAu}$ is seen if compared at the same multiplicity. Similar to the CMS observation, this seems to indicate that the previously observed 
$\Delta \gamma_{112}$ in peripheral Au+Au collisions are entirely dominated by backgrounds. However, note that a key difference between
the CMS and STAR analyses is the $\eta$ gap imposed. The STAR analysis
implements a much smaller $\eta$ gap between particle $\alpha$,$\beta$ and the event plane, and thus likely includes additional contribution of short-range correlations. Indeed, when different $\eta$ gaps are introduced, the results
could vary by a factor of 2 in p+Au and d+Au~\cite{STAR6}. The multiplicity range
covered by STAR and CMS analyses is also quite different, which may result in
different long-range nonflow correlation backgrounds. Future program with small systems and upgraded detectors at RHIC and the LHC will help provide a
more meaningful and consistent comparison between difference energies.

\subsubsection{Correlators relative to the third-order event plane}
The $\gamma_{112}$ measurements are usually implemented with the second-order event plane, which is approximately perpendicular to the magnetic field direction. A new correlator, $\gamma_{123}$~\cite{CMS2}, with respect to the third-order event plane, was motivated for the background study,
\begin{equation}
    \gamma_{123} \equiv \langle \langle \cos(\phi_\alpha + 2\phi_\beta -3{\rm \Psi_{3}}) \rangle\rangle.
\end{equation}
As the third-order event plane is entirely driven by participant fluctuations, it is largely decoupled from the reaction plane or the direction of the magnetic field. 
Therefore, to a good approximation, the 
$\Delta\gamma_{123}$ correlator captures pure background 
contributions that are proportional to $v_3 \cdot \Delta\delta$ 
(following a similar derivation to Eq.~\ref{eq:11}). 
Similar to Eq.~\ref{kappa112}, a normalized quantity can be
defined~\cite{CMS2}:
\begin{equation}
    \kappa_{123} \equiv \frac{\Delta \gamma_{123}}{v_3 \cdot \Delta \delta},
\end{equation}
which characterizes purely the backgrounds.

\begin{figure}[h]
\includegraphics[width=2.5in]{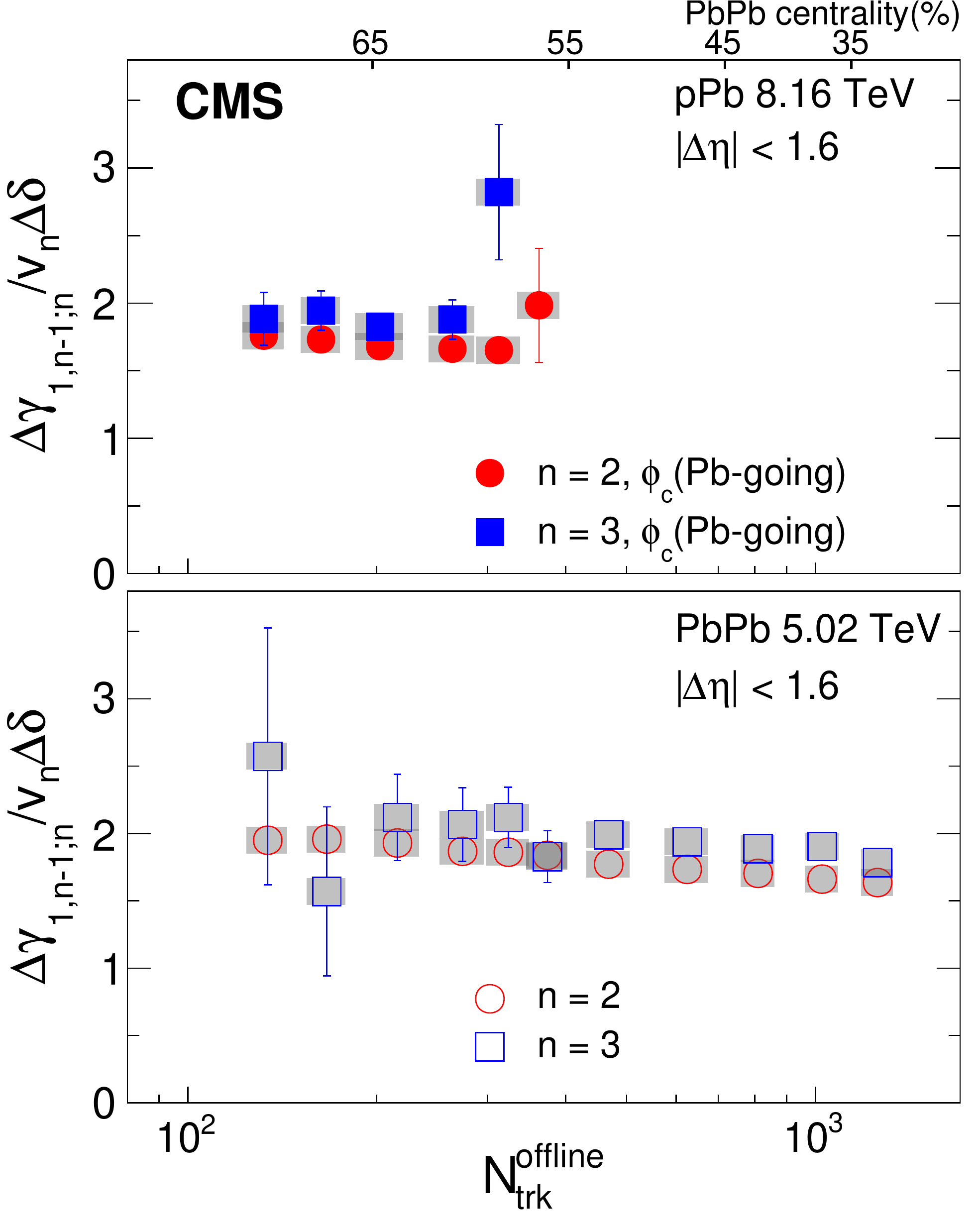}
\caption{$\kappa_{112}$ and $\kappa_{123}$ measured with the participant plane in p+Pb collisions for the Pb-going direction at 8.16 TeV (upper) and Pb+Pb collisions
at 5.02 TeV (lower)~\cite{CMS1,CMS2}.}
\label{fig4}
\end{figure}

It is intuitive to use $\kappa_{123}$ as a data-driven background estimate for $\kappa_{112}$. Indeed, both CMS data of Pb+Pb collisions at 5.02 TeV~\cite{CMS2} (illustrated in the lower panel of Fig.~\ref{fig4}) and STAR preliminary data of Au+Au collisions at 200 GeV~\cite{Kong} show that $\kappa_{112}$ and $\kappa_{123}$ are close to each other in most centrality intervals studied.
This indicates little fraction, if any, of the CME contribution 
to the measured $\kappa_{112}$. The expectation of similar 
$\kappa_{112}$ and $\kappa_{123}$ values in a pure 
background environment is further validated by the CMS p+Pb 
data (entirely dominated by backgrounds) as shown in the upper panel of
Fig.~\ref{fig4}, again suggesting little room for any CME contribution in Pb+Pb data at LHC energies.

In model calculations without any CME signal, AMPT suggests that
$\kappa_{112}$ and $\kappa_{123}$ are not identical in Au+Au collisions at 
200 GeV, with the latter higher than the former by about 50\% (or a 
ratio of $3/2$)~\cite{Subikash}. This indicates that higher-harmonic 
anisotropy is more ``damped" in the system evolution, and $\kappa_{123}$ is 
at best a qualitative estimate for $\kappa_{112}$, which may depend heavily 
on properties of system evolution. This may not be surprising, based on previous derivations, where $\kappa_{112}$ and $\kappa_{123}$ are not exactly identical. Looking closely at the CMS data
in both Pb+Pb and p+Pb in Fig.~\ref{fig4}, $\kappa_{123}$ is indeed
generally above $\kappa_{112}$. More work is still needed to better understand and/or redefine the $\kappa_{123}$ to draw
quantitative conclusion on the possible CME contribution to $\kappa_{112}$.

Another derivative of the $\gamma_{112}$
correlator and its normalized quantity are~\cite{Subikash}
\begin{eqnarray}
\gamma_{132} &\equiv& \langle \langle \cos(\phi_\alpha - 3\phi_\beta + 2{\rm \Psi_{2}}) \rangle\rangle \nonumber\\
\kappa_{132} &\equiv& \frac{\Delta \gamma_{132}}{v_2 \cdot \Delta \delta}.
\end{eqnarray}
Although $\gamma_{132}$ uses the second-order event plane, it is similar to $\gamma_{123}$
in the sense that it is linked to the CME only via a $v_2 \cdot \Delta\delta$ term. 
The AMPT simulations~\cite{Subikash} show that $\kappa_{132}$ is close to unity in most cases, providing another data-driven gauge
of the background baseline.

\subsubsection{Event shape engineering}
The event-shape engineering
technique was proposed to quantitatively remove backgrounds related to long-range
collective (or flow) correlations in a data-driven way~\cite{ESE1,ESE2}.
The idea is based on the expectation that the CME signal is independent
of $v_2$, while the dominant background is proportional to $v_2$, as supported by $\kappa_{112}$ and $\kappa_{123}$ data in p+Pb:
\begin{equation}
    \Delta \gamma_{112} = \kappa_{112}^{\rm BKG} \cdot v_2 \cdot \Delta \delta + \Delta \gamma_{112}^{\rm CME}. 
\label{eq:ese}    
\end{equation}
If extrapolating Eq.~\ref{eq:ese} to the scenario of $v_2 = 0$, it will then arrive at $\Delta \gamma_{112}|_{v_2=0} = \Delta\gamma_{112}^{\rm CME}$. Note that some dependence of the CME signal on $v_2$ is expected when $v_2$ becomes very small and event plane resolution is poor. This effect is studied by Monte Carlo models, as will be mentioned later.

The ``standard" procedure of event-shape engineering is to keep the following three types of particles independent of each other in an event:
(A) the particles that are used to engineer the event shape, (B) the particles of interest ($\alpha,\beta$), and (C) the particles that reconstruct the event plane ($\Psi_{\rm EP}$). In other words, they should come from three different sub-events. In practice, the flow vector of sub-event A, $\overrightarrow{q} = (q_x^{\rm A},q_y^{\rm A})$, controls the event shape:
\begin{eqnarray}
q_x^{\rm A} &=& \frac{1}{\sqrt{N}} \sum_i^N \cos(2\phi_i^{\rm A}) \label{qx}  \\
q_y^{\rm A} &=& \frac{1}{\sqrt{N}} \sum_i^N \sin(2\phi_i^{\rm A}), \label{qy}
\end{eqnarray}
where the magnitude of $\overrightarrow{q}$ is equivalent to 
$v_2$ but contains effects of statistical fluctuations.
For different $q^{\rm A}$ bins, $v_2^{\rm B}$ and $\Delta\gamma_{112}^{\rm B}$ are calculated for particles in sub-event B, with the event plane estimated from sub-event C.
Then $\Delta\gamma_{112}^{\rm B}$ is plotted as a function of $v_2^{\rm B}$, and the extrapolation of $\Delta\gamma_{112}^{\rm B}$ to the value at $v_2^{\rm B}=0$ yields the true CME contribution with no flow-related backgrounds. Although $q^{\rm A}$ and $q^{\rm B}$ are linearly correlated on average,
there is a spread between them on an event-by-event basis, arising from statistical fluctuations. Therefore, even the lowest $q^{\rm A}$ bin close to zero could correspond to a sizable $v_2^{\rm B}$.  Systematic uncertainties 
and model dependence could be introduced when $\Delta\gamma_{112}^{\rm B}$ is extrapolated over a wide unmeasured $v_2^{\rm B}$ region.

\begin{figure}[th]
\includegraphics[width=\textwidth]{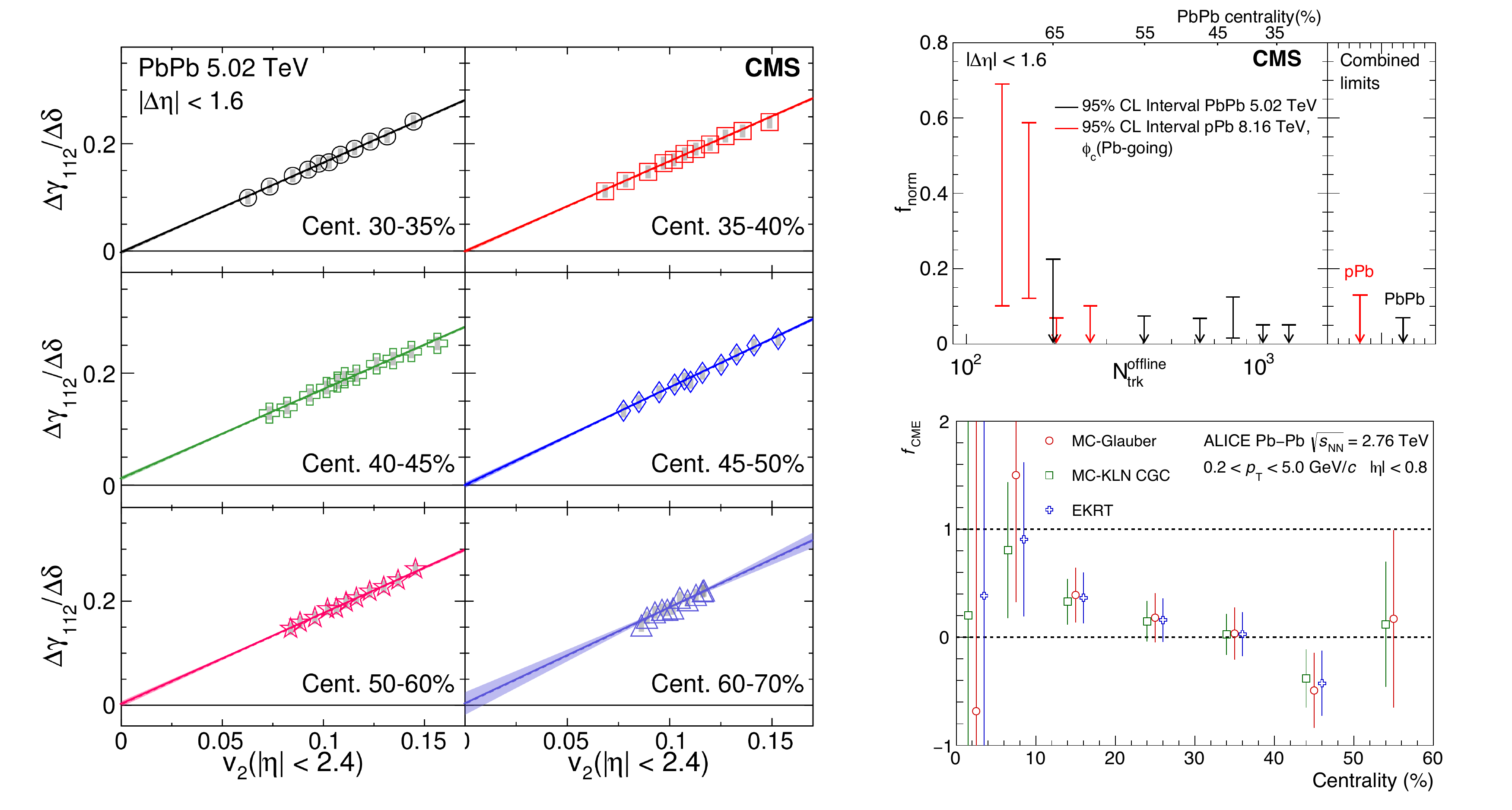}
\caption{Left: ratios of $\Delta\gamma_{112}$ to $\Delta\delta$ as a function of $v_2$ 
for different centrality classes in Pb+Pb collisions at 5.02 TeV. Right: upper limits on
fractions of possible CME signals in $\Delta\gamma_{112}$.}
\label{fig:ese_alicecms}
\end{figure}

Figure~\ref{fig:ese_alicecms} demonstrates the application of this 
event-shape engineering approach by the CMS collaboration 
in Pb+Pb collisions at 5.02 TeV for various centrality ranges~\cite{CMS2}. A ratio
of $\Delta\gamma_{112}$ to $\Delta\delta$ is taken and plotted as a function of $v_2$ to eliminate a weak dependence of $\Delta\delta$ on
$v_2$ (especially for peripheral events). A linear fit to data and extrapolation to $v_2=0$
results in a vertical intercept well consistent with zero, indicating little or no CME signal.
Under the assumption of linear extrapolation, an upper
limit on the fraction of possible CME signals in $\Delta\gamma_{112}$ can be extracted, 
which is less than 7\% at 95\% confidence level (C.L.) in Pb+Pb collisions from the CMS data, shown in Fig.~\ref{fig:ese_alicecms} (top right). 
As mentioned earlier, a large fraction of $v_2$ region towards zero $v_2$ is lack of data points and relies entirely on extrapolation. The
ALICE collaboration took into account possible $v_2$ dependence
of the CME signal as $v_2$ approaches to zero based on several Monte Carlo initial-state models, and also obtained fractions of
residual CME signals in $\Delta\gamma_{112}$, shown in Fig.~\ref{fig:ese_alicecms} (bottom right). The conclusion of CMS and ALICE measurements are consistently holding
that at the LHC energies, the observed $\Delta\gamma_{112}$ correlator is consistent with 100\% flow-driven background contributions. The CME signal contribution to the $\Delta\gamma_{112}$ correlator, if indeed present, has an upper limit of only a few \%.

One way to avoid the long extrapolation in $v_2$ is to 
sacrifice the independence between sub-events A and B.
When particles of interest are used to define $q$, the lowest $q$ bin naturally corresponds to a $v_2$ value very close to zero, and the extrapolation is technically much more reliable. The caveat on this approach is that only the ``apparent" flow (including statistical fluctuations) is under control.
It is still possible that a resonance parent has a finite $v_2$, and its decay daughters have zero contribution to $q$. In this case, even at zero $q$ or $v_2$, there exists a fake CME signal in $\Delta\gamma_{112}$.
This approach has been tested with a background-only AMPT model. The disappearance of background is demonstrated when $\Delta\gamma_{112}$ is extrapolated to zero $q$~\cite{Fufang}. Another caveat on the event-shape engineering in general
is that in reality the CME signal and flow magnitude could have an intrinsic correlation, e.g., related to the centrality or impact parameter dependence.
In that case, the projection to $v_2=0$ could lead to an over-subtraction of the background. 
Model studies on the centrality dependence of the magnetic field and
initial-state eccentricity will help understand  this effect.

\subsubsection{Other developments}
Derivatives of the $\gamma_{112}$ observable with respect to different types of event planes, i.e., the participant plane ($\Psi_{\rm PP}$) and the spectator plane ($\Psi_{\rm SP}$), were also proposed. Here, the idea is that the CME signal is most correlated with $\Psi_{\rm SP}$, while the flow-driven backgrounds are largely correlated with $\Psi_{\rm PP}$. Making use of the decorrelation between $\Psi_{\rm SP}$ and $\Psi_{\rm PP}$,  $\Delta \gamma_{112}$ can be decomposed into two components: $\Delta \gamma_{112}^{\rm flow}$ and $\Delta \gamma_{112}^{\rm CME}$, following the relations below~\cite{RP-PP}:
\begin{eqnarray}
\Delta \gamma_{112}\{\rm PP\} &=& \Delta \gamma_{112}^{\rm flow}\{\rm PP\} + \Delta \gamma_{112}^{\rm CME}\{\rm PP\},  \\ 
\Delta \gamma_{112}\{\rm SP\}
&=& \Delta \gamma_{112}^{\rm flow}\{\rm SP\} + \Delta \gamma_{112}^{\rm CME}\{\rm SP\},  \\
&=& a\cdot \Delta \gamma_{112}^{\rm flow}\{\rm PP\}  + \Delta \gamma_{112}^{\rm CME}\{\rm PP\} /{\it  a} ,
\end{eqnarray}
where $a = \langle \cos(2\Psi_{\rm PP} - 2\Psi_{\rm SP}) \rangle$, denoting the decorrelation between the two event planes. By eliminating $\Delta \gamma_{112}^{\rm flow}\{\rm PP\}$ from the equations, the CME signal can be extracted,
\begin{equation}
    \Delta \gamma_{112}^{\rm CME}\{\rm PP\} = (\Delta \gamma_{112}\{\rm SP\} - {\it  a}\cdot\Delta \gamma_{112}\{\rm PP\})/(1/{\it  a} -{\it  a}).
    \label{PP}
\end{equation} 
In fact, the $\gamma$ correlator data with respect to $\Psi_{\rm PP}$ and $\Psi_{\rm SP}$ are already presented in Fig.~\ref{fig1} (corresponding to $\Psi_{2}$ and $\Psi_{1}$), but higher statistics are needed for a firm conclusion. 

There are also other method developments involving differential measurements of the $\gamma$ correlators, such as those as functions of the particle pair's relative pseudorapidity ($\Delta\eta$)~\cite{Prithwish} and invariant mass ($m_{\rm inv}$)~\cite{Mass}. The former approach assumes that the CME-induced correlations should not be of short ranges in rapidity, and hence the decomposition of $\gamma(\Delta\eta)$ into several Gaussian distributions could separate the contributions of different physics origins, or at least exclude the short-range correlations. However, the typical correlation length of the CME-induced correlations is still elusive on the theoretical side, and for a clear interpretation, the $\gamma_{112}(\Delta\eta)$ results need to be compared with other correlators that are dominated by backgrounds, such as $\gamma_{123}(\Delta\eta)$ and $\gamma_{132}(\Delta\eta)$. Data from CMS in p+Pb and Pb+Pb collisions do not indicate any obvious difference in $\Delta\eta$ dependence of $\gamma_{112}$ and $\gamma_{123}$ correlators. The latter approach focuses on the backgrounds due to resonance decays, and attempts to remove such contributions by rejecting particle pairs with small $m_{\rm inv}$. Again, the theoretical guidance is needed on the $m_{\rm inv}$ dependence of the CME signal, the lack of which hinders a definite conclusion.

\subsection{Outlook}
Along the line of disentangling the possible CME signal and flow-driven backgrounds via data-driven approaches, 
comparison between U+U and Au+Au collisions as a function of centrality is another promising way of obtaining new insights. A uranium nucleus have 13 more protons than a gold nucleus, which
in turn causes stronger magnetic fields in U+U than Au+Au collisions at the same number of participating nucleons ($N_{\rm part}$). The difference in the magnetic field is compensated by the difference in ellipticity at lower
$N_{\rm part}$, but becomes overwhelming towards very central events with higher $N_{\rm part}$. This direction is being pursued at RHIC.

\begin{figure}[h]
\includegraphics[width=0.9\textwidth]{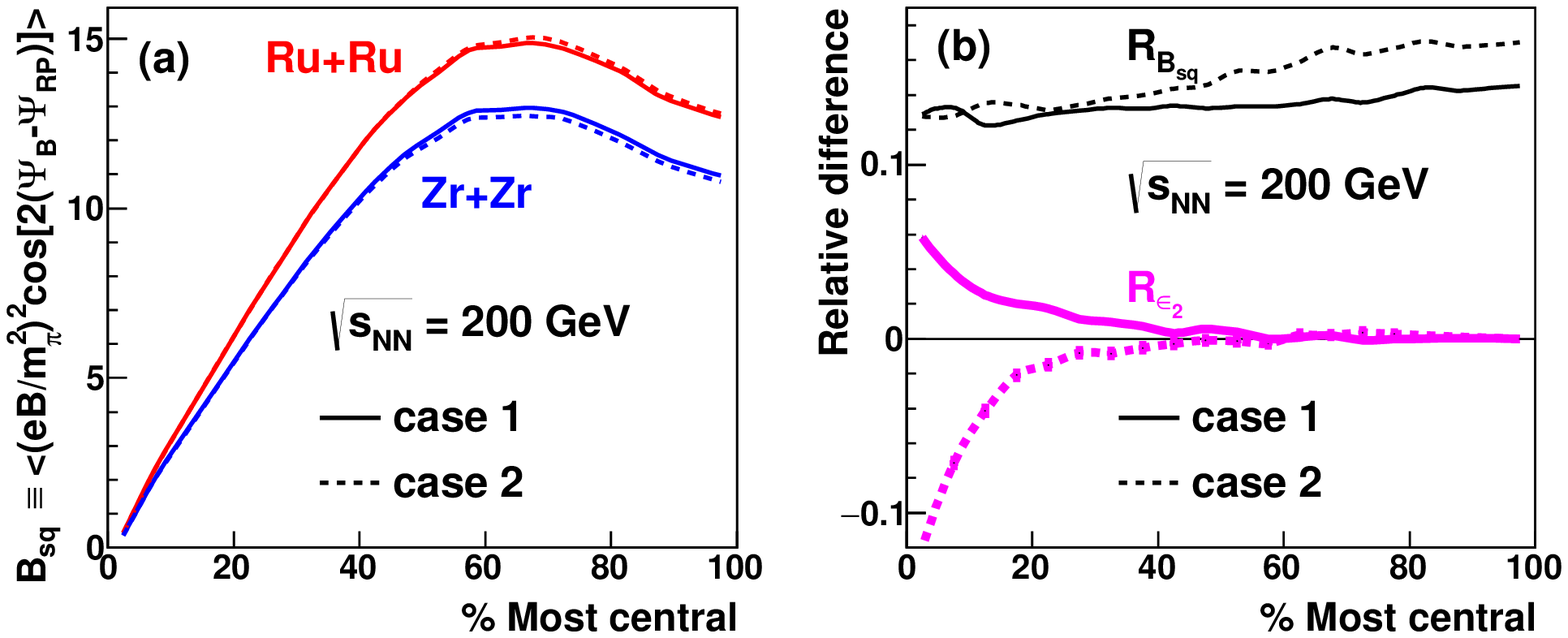}
\caption{Theoretical calculation of the initial magnetic field squared with correction from azimuthal fluctuation (a) and their relative difference (b) versus centrality for Ru+Ru and Zr+Zr collisions at 200 GeV~\cite{isobar}. Also shown is the relative difference in
initial eccentricity. The solid (dashed) lines correspond to the deformity parameter set of case 1 (case 2).
}
\label{fig5a}
\end{figure}

Following this direction of fixing $v_2$-driven backgrounds, collisions of isobaric nuclei, such as $^{96}_{44}$Ru and $^{96}_{40}$Zr, have been proposed~\cite{UU_theory2}.
The Ru+Ru and Zr+Zr collisions at the same beam energy are almost identical in terms of hadronic particle 
production,
but their initial magnetic fields differ because of the charge difference between the Ru and Zr nuclei.
Figure~\ref{fig5a} shows that the relative difference in magnetic field magnitudes between Ru+Ru and Zr+Zr collisions is
around $13\%$ for central events, and increases to 
$15\%-18\%$ for peripheral events~\cite{isobar}.
The systematic uncertainty in peripheral collisions is related to our incomplete knowledge of the deformity ($\beta_2$) of Ru and Zr nuclei~\cite{e-A1,e-A2,ModelFit}. There is a little relative difference in eccentricity (that drives $v_2$), but much smaller than that in the magnitude field. AMPT simulations have shown that the relative difference in the CME signal embedded in the model between the two isobars is robust and can survive the final-state interactions~\cite{isobar2}.

The isobaric program was carried out at RHIC in 2018,
where about 3 billion minimum bias events for each of the two isobaric systems have been collected by the STAR experiment. The data are being analyzed.
In a simple scenario of identical $v_2$ between Ru+Ru and Zr+Zr systems, there are three signatures that would indicate a CME signal~\cite{Jinfeng}: (1) $\Delta\gamma_{112}^{\rm Ru+Ru} > \Delta\gamma_{112}^{\rm Zr+Zr}$; (2) $\Delta\delta_{112}^{\rm Ru+Ru} < \Delta\delta_{112}^{\rm Zr+Zr}$; (3) $(\Delta\gamma_{112}^{\rm Ru+Ru} - \Delta\gamma_{112}^{\rm Zr+Zr}) / (\Delta\delta_{112}^{\rm Ru+Ru}-\Delta\delta_{112}^{\rm Zr+Zr}) = \langle \cos[2(\Psi_{RP} - \Psi_{\rm EP})] \rangle $.
If the CME contributes to more than 20\% of the $\Delta\gamma_{112}$ signal in the 20--60\% centrality range, a more than $7\sigma$ difference between Ru+Ru and Zr+Zr data is expected.
Taking into account possible difference in $v_2$ between the two isobaric systems, the $\kappa_{112}$ and $\kappa_{123}$ observable can be invoked to mitigate the effect for a fair comparison.
If no difference is seen between Ru+Ru and Zr+Zr within experimental uncertainties, upper limits on the
CME contribution to $\Delta\gamma_{112}$ and other CME-motivated observable can be obtained. However,
it should be also noted that there are other effects 
that may shadow the charge difference between the two isobars. For example, if Zr has a thicker neutron skin than Ru, the initial magnetic field in Zr+Zr is then stronger than originally estimated, and the difference in the magnetic field between Zr+Zr and Ru+Ru will diminish~\cite{skin1,skin2}.

The RHIC BES-II program has been ongoing from 2018 to 2022, with the main goal of searching for a critical endpoint of the QCD phase diagram. It will significantly
expand the data sets collected during the BES-I phase for a variety of collision energy.
The projected statistical uncertainties on the $\kappa_{112}$ measurement is displayed with the shaded band in Fig.~\ref{fig3}, which are nearly invisible.
These future results and in comparison with realistic model calculations, such as AMPT and anomalous-viscous fluid dynamics (AVFD) model~\cite{AVFD}, 
will draw a more definitive conclusion on the possible disappearance of CME signal in Au+Au collisions at low beam energies. 
The newly installed Event Plane Detector (EPD) in the STAR experiment introduces a sizable $\eta$ gap between the particles of interest and the event plane, and will help suppress the nonflow effects, which are particularly sizable for low multiplicity events at lower beam energies. For some beam energies, the $\eta$ coverage of the EPD could span into the beam rapidity  to capture spectator nucleons. This will facilitate the comparison between the $\Delta\gamma_{112}\{\rm SP\}$ and $\Delta\gamma_{112}\{\rm PP\}$ correlators.
\section{CMW searches in nuclear collisions}
\label{sec:CMW}

In this section, searches for another anomalous chiral effect, chiral magnetic wave (CMW), in nuclear collisions is reviewed.
Methodologies employed to search for the CMW signal are outlined, followed by reviews of experimental results at RHIC~\cite{CMW_STAR} and the LHC~\cite{CMW_ALICE,CMW_CMS} in both large and small systems. Background contributions to the CMW searches are discussed. The section ends with a future outlook.

\subsection{Methodologies}

In heavy-ion collisions, the CMW-induced electric quadrupole evolves with hydrodynamic expansion of the QGP, and results in a charge-dependent elliptic flow ($v_2$).
Taking pions as an example, the $v_2$ values for $\pi^{+}$ and $\pi^{-}$ are expected
to be identical from hydrodynamic flow alone.
On top of this baseline $v_2^{\rm
  base}(\pi^\pm)$, the CMW will introduce  an additional contribution~\cite{CMW}
\begin{equation}
v_2(\pi^\pm) = v_2^{\rm base}(\pi^\pm) \mp (\frac{q_e}{\bar \rho_e})A_{\rm ch},
\label{eq:v2_slope}
\end{equation}
where $q_e$, ${\bar \rho_e}$ and $A_{\rm ch} = (N_+ - N_-)/(N_+ + N_-)$ are the quadrupole moment,
the net charge density and the final-state charge asymmetry of a collision event, respectively.
With $\langle A_{\rm ch} \rangle$ always positive,
the $A_{\rm ch}$-integrated $v_2$ of $\pi^-$ ($\pi^+$) should be above (below)
the baseline because of the CMW, leading to a splitting between $v_2(\pi^+)$ and $v_2(\pi^-)$. For $A_{\rm ch}$-integrated $v_2$, it has been proposed that the baseline $v_2$ for $\pi^+$ and $\pi^-$ may be modified by other physics mechanisms~\cite{Dunlop:2011cf,Xu:2012gf} unrelated to the CMW. Therefore, the most unambiguous way to search for the CMW signal in nuclear collisions is 
to measure the full $A_{\rm ch}$ dependence of pion $v_2$, or the slope of $\Delta v_2$ ($v_2$ difference between $\pi^-$ and $\pi^+$) as a function of $A_{\rm ch}$.
In the following discussions, the $r_2$ parameter is used to represent the slope of $\Delta v_2(A_{\rm ch})$, where a positive $r_2$ indicates a possible CMW signal. 

In the method above, a correction to the observed $A_{\rm ch}$ for the finite detector efficiency is required. An alternative approach of a three-particle correlator
that is less dependent on the efficiency correction was proposed~\cite{Voloshin_Belmont},
\begin{equation}
\langle \langle \cos[n(\phi_1-\phi_2)]q_3 \rangle \rangle = \langle \cos[n(\phi_1-\phi_2)]q_3 \rangle - \langle \cos[n(\phi_1-\phi_2)] \rangle \langle q_3 \rangle.
\label{DAQ}
\end{equation}
Here $\phi_1$ and $\phi_2$ are the azimuthal angles of particles 1 and 2 ($\pi^\pm$), and $q_3$ is the charge ($\pm1$) of particle 3.
The single brackets represent the average over particles and events, and the double bracket denotes the cumulant.
Without charge-dependent correlations, this correlator should be equal to zero. A positive difference in this correlator between $\pi^-$ and $\pi^+$ will signify the CMW contribution.

Similar to searches for the CME, more alternative methods have been proposed that may
provide different sensitivity to the CMW signal and backgrounds. For example,
in analogy to the multiparticle charge-dependent correlator for the 
CME search~\cite{PHENIX1,PHENIX2,Roy}, a novel correlator 
has been recently proposed in search of the CMW-induced electric quadrupole~\cite{CMW_AMPT}.
This new method does not require $A_{\rm ch}$ either.
Using AMPT model calculations, this approach has shown a potential in improving the methodology of the CMW search.

In reviewing experimental results below, we focus on the slope parameter $r_2$ of $\Delta v_2(A_{\rm ch})$ and its derivatives, which is the most widely studied observable.

\subsection{Results in A+A collisions: evidence for the CMW}

\begin{figure}[th]
\includegraphics[width=0.8\textwidth]{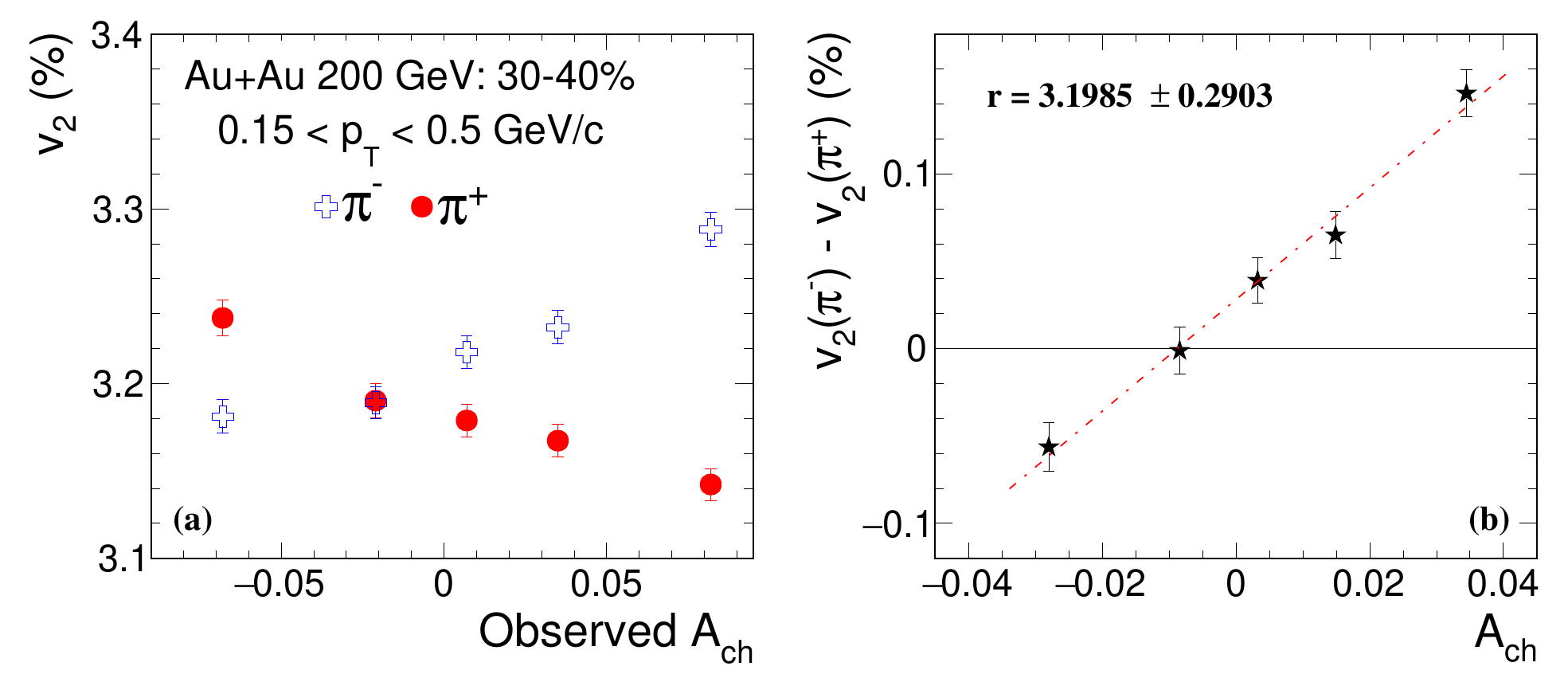}
\caption{Left: pion $v_2$ as a function of observed
charge asymmetry. Right: $v_2$ difference between $\pi^{-}$ and $\pi^{+}$ 
as a function of charge asymmetry after corrected for the detector efficiency, 
for 30--40\% centrality Au+Au collisions at
200 GeV.~\cite{CMW_STAR}
}
\label{fig:cmw_star}
\end{figure}

First evidence for the CMW has been reported by STAR in the measurement of charge-dependent
pion $v_2$ as a function of $A_{\rm ch}$ in Au+Au collisions at 200 GeV, shown in Fig.~\ref{fig:cmw_star} (left)~\cite{CMW_STAR}. A significant splitting between $\pi^{+}$ and $\pi^{-}$ $v_2$ data is observed, which has an
approximately linear dependence on $A_{\rm ch}$. The $\Delta v_2 (A_{\rm ch})$ result is also shown in Fig.~\ref{fig:cmw_star} (right), where the slope parameter $r_2$ is extracted
by a linear fit. These observations are consistent with the expectation of the CMW.

\begin{figure}[th]
\includegraphics[width=\textwidth]{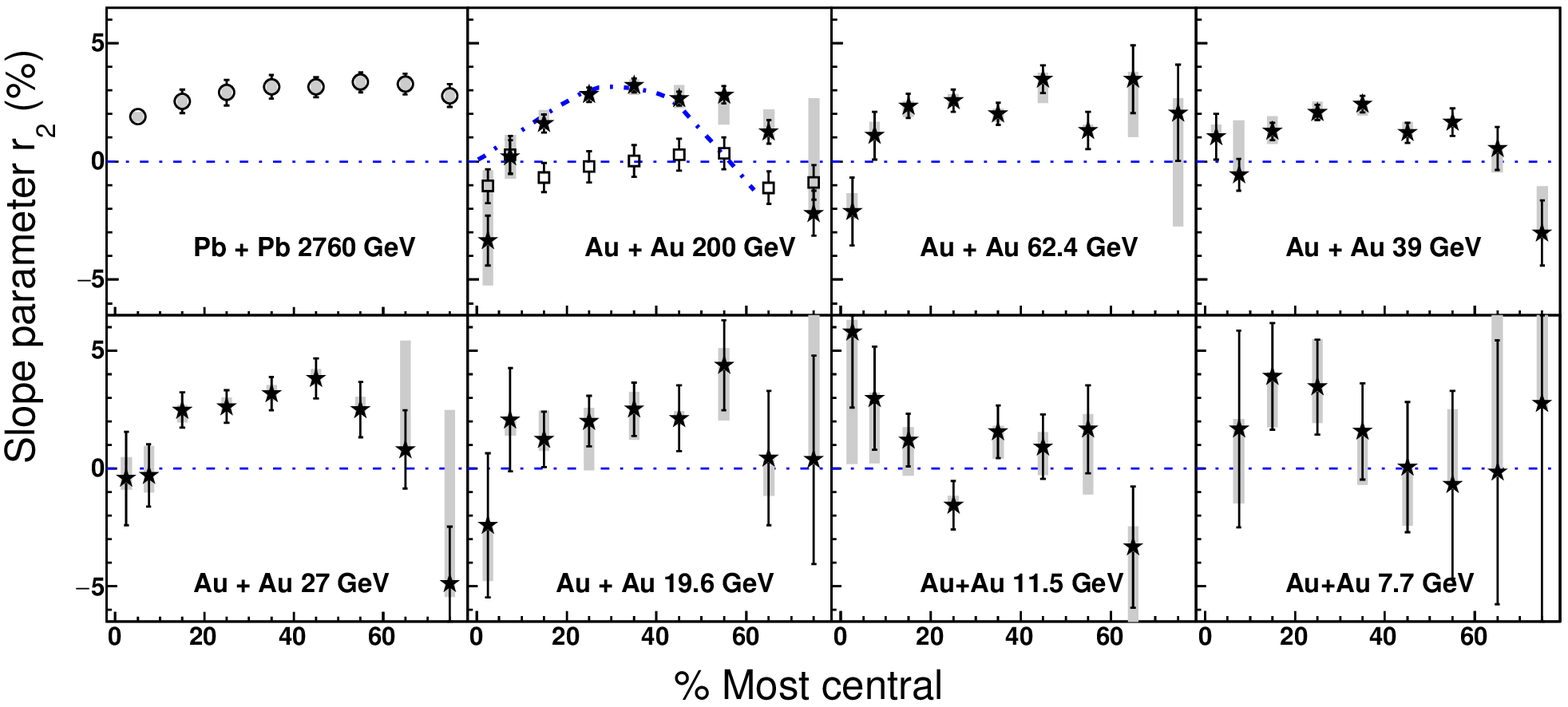}
\caption{The slope parameter ($r_2$) as a function of centrality for Pb+Pb collisions at 2.76 TeV~\cite{CMW_ALICE} and for Au+Au collisions at 7.7-200 GeV~\cite{CMW_STAR}. The grey bands represent systematic uncertainties. In comparison with data of 200 GeV Au+Au, UrQMD calculations~\cite{UrQMD} (open square) and a CMW model calculation with a magnetic field duration time of 5 fm/$c$~\cite{CMW_calc_Burnier} (dashed-dotted line) are also shown.
}
\label{fig6}
\end{figure}

The measured slope parameter ($r_2$) is presented in Fig.~\ref{fig6} as a function of centrality for
Pb+Pb collisions at 2.76 TeV~\cite{CMW_ALICE} and for Au+Au collisions from 7.7 to 200 GeV~\cite{CMW_STAR}.
A universal rise-and-fall trend is observed in the centrality dependence of $r_2$ for most beam energies
except for 11.5 and 7.7 GeV, where the $r_2$ slopes are consistent with zero, although statistical uncertainties are still large. This trend is in line with the CMW expectation as it approximately follows how the magnitude of the magnetic field is expected to evolve with centrality.
The comparison between the STAR data for 200 GeV Au+Au and the ALICE data for 2.76 TeV Pb+Pb
reveals a striking similarity, especially considering the many differences between the two measurements such as collision energies, multiplicities, and kinematic acceptance:
particles of interest in the STAR data are charged pions with $0.15<p_T<0.5$ GeV/$c$ and $|\eta|<1$,
while those in the ALICE data are unidentified hadrons with $0.2<p_T<5$ GeV/$c$ and $|\eta|<0.8$.
Note that in most central and peripheral events, the $r_2$ data at 200 GeV are consistent with zero within uncertainties or even negative, whereas it still remains significantly positive at 2.76 TeV.
This may indicate an interplay of different physics mechanisms between RHIC and LHC energies.

The three-particle correlator (as defined in Eq.~\ref{DAQ}) for the $2^{\rm nd}$ and $3^{\rm rd}$ harmonics has been measured by ALICE as a function of centrality in Pb+Pb collisions at 2.76 TeV~\cite{CMW_ALICE}.
The ordering between negative and positive particles of interest also supports the CME picture in noncentral collisions.
The correlation strength substantially increases in more peripheral collisions,
which may go beyond the pure CMW interpretation.
Not surprisingly, background sources (e.g., the LCC effect) could also
contribute to the $r_2$ measurement, as for the 
$\gamma$ correlator in the CME search.

\subsection{Approaches to disentangle the signal vs. background}

There are two major types of physical backgrounds proposed that may contribute to the $r_2$ measurement: isospin chemical potential ($\mu_I$)~\cite{isospin}, namely imbalance between numbers of up and down quarks, and local charge conservation (LCC) effect~\cite{Bzdak2013}, similar to that for the CME search.

A hydrodynamic study~\cite{isospin} incorporating finite initial $\mu_I$ 
suggests that a simple viscous transport of charges, combined with certain 
initial conditions, will lead to a sizable $v_2$ splitting for charged pions. 
According to analytical calculations of the anisotropic Gubser flow, the 
$\Delta v_2$ for pions is proportional to both the shear viscosity and $\mu_I$, and $\mu_I$ is linearly related to $A_{\rm ch}$. This model
further predicts a negative $r_2$ for charged kaons with larger magnitudes than 
the pion $r_2$, since $\mu_I$ and strangeness chemical potential ($\mu_S$)
are different for kaons and pions. This idea does not seem to be supported by 
preliminary STAR data showing that the kaon $r_2$ is consistent with the pion $r_2$ 
within uncertainties for various collision energies at RHIC~\cite{Qi-ye2,Qi-ye3}.

The LCC mechanism is also able to qualitatively explain the positive $r_2$ 
observed from data, when convoluted with the characteristic dependence of 
$v_2$ on $\eta$ and $p_T$ in a limited detector acceptance~\cite{Bzdak2013}. Similar to the CME background study, 
this effect to the CMW is demonstrated with a model of ``flowing'' clusters that 
locally conserve charges when decaying into final-state particles, e.g., a pair of particles with opposite charges. Such a pair could cause a
non-zero $A_{\rm ch}$, if one of the particles falls outside 
the detector acceptance. If this process preferentially occurs 
in a phase space with smaller $v_2$, such as a lower-$p_T$ or higher-$\eta$ 
region, a positive $r_2$ can then be induced, no matter the missing particle 
is positive- or negative-charged. A first estimate of the LCC contribution within a simplified model,
however, appears to underestimate the $r_2$ observed in the STAR measurements 
in Au+Au at 200 GeV by an order of magnitude~\cite{CMW_STAR}.
The LCC scenario also predicts that a narrower $\eta$
coverage may lead to a stronger $r_2$, since it is more likely that one
particle from a charge-conserved cluster escapes the detector.
In a preliminary STAR result where the $\eta$ coverage is reduced to half,
$r_2$ does not seem to display a significant variation within experimental uncertainties~\cite{Qi-ye3}. Another prediction by the LCC scenario is that the $r_2$ slope is proportional to the baseline $v_2$. This motivated a new noramlized $r_2$ observable, $r_2^{\rm norm}=r_2/v_2$, by the CMS collaboration~\cite{CMW_CMS}.

Transport Monte Carlo models have been used to study contributions from backgrounds or conventional physics.
The slope parameters extracted from UrQMD 
calculations for Au+Au collisions at 200 GeV~\cite{UrQMD}, 
shown in Fig.~\ref{fig6}, are consistent with zero for the 
10--70\% centrality range, as opposed to the positive signal observed in the real data.
On the other hand, the simplified CMW calculations with a magnetic field 
duration time of 5 fm/$c$ demonstrate a centrality dependence of $r_2$ 
similar to the data~\cite{CMW_calc_Burnier}. AMPT calculations also show that a positive
$r_2$ caused by the electric quadrupole can survive the final-state
interactions~\cite{CMW_AMPT}.
A quantitative comparison between data and theoretical calculations still requires further work. 
In particular, a model that incorporates both the CMW signal and background contributions (
such as the AVFD model~\cite{AVFD}) will help facilitate the comparison with data.

The transported-quark model~\cite{Dunlop:2011cf} argues that at lower beam
energies, the $A_{\rm ch}$-integrated $v_2$ difference between particles and
anti-particles can be explained by the effect of quark transport
from projectile nucleons to mid-rapidity. The assumption is that
quark coalescence mechanism still holds at low energies, and
the $v_2$ of transported quarks is larger than that of produced quarks. 
The model, however, suggests a negative slope for $v_2(\pi^-) - v_2(\pi^+)$
as a function of $A_{\rm ch}$~\cite{Campbell}, which is
opposite to the data for most centrality intervals.

Data-driven approaches to constrain the CMW background contributions to the $r_2$ measurement are discussed below,
following the same strategy as for the CME search to vary the signal (background) while keeping the background (signal) fixed, using small-system collisions and higher-order flow coefficients.

\subsubsection{Results in small systems}

Following the same idea as for the CME search, measurements of charge-dependent
$v_2$ and slope parameter $r_2$ in small p+A systems can provide a baseline of 
pure background contributions, because of decorrelation between the event plane and the
magnetic field direction. Figure~\ref{fig7} displays the CMS measurements of normalized $r_2$ for charged hadrons in p+Pb and Pb+Pb collisions at 5.02 TeV~\cite{CMW_CMS}. 
The $r_{2}^{\rm norm}$ values are comparable between p+Pb and Pb+Pb collisions (with p+Pb
data even larger), with little dependence on event multiplicity or centrality.
Similar to the conclusion drawn for the CME, these results again suggest
that the $r_2$ slopes observed in peripheral Pb+Pb collisions at 5.02 TeV are likely to be dominated by background contributions unrelated to the CMW. 

Preliminary STAR results show that in minimum bias p+Au and d+Au collisions at 200 GeV, 
the $r_2$ value is consistent with zero within current uncertainties. This would support
the observation of a CMW signal in Au+Au collisions at 200 GeV~\cite{Qi-ye2,Qi-ye3}. As discussed earlier for the CME, an energy dependence of the CMW signal is possible due to
much shorter lifetime of magnetic field at higher energies when two ions pass by much faster.
However, higher precision small-system data at RHIC, especially covering a wide range of
multiplicity as for the LHC data, are still needed to draw a definitive conclusion.

\begin{figure}[th]
\begin{minipage}[c]{0.48\textwidth}
\includegraphics[width=\textwidth]{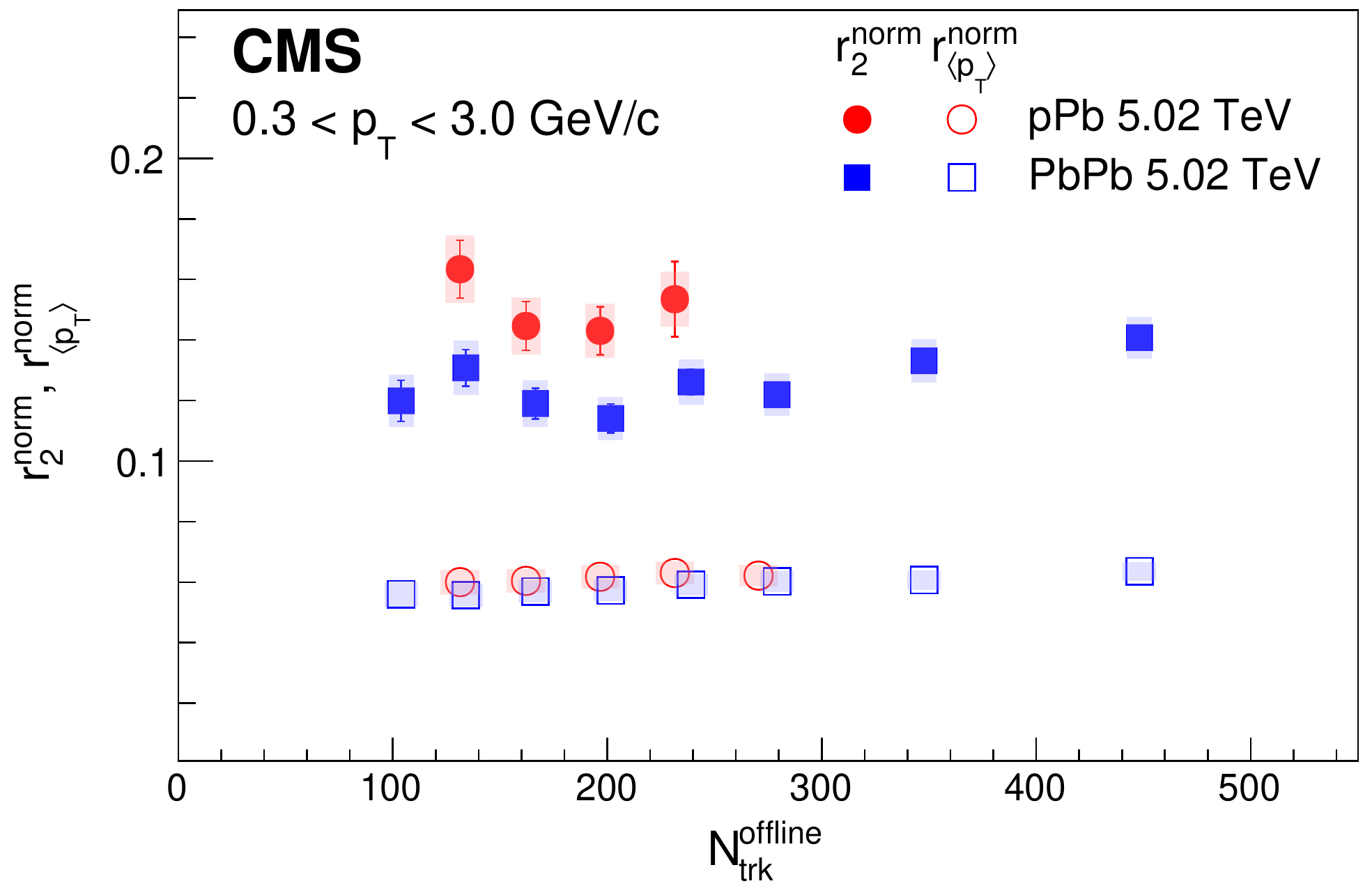}
\end{minipage}
\hspace{0.02\textwidth}
\begin{minipage}[c]{0.49\textwidth}
\vspace{-0.2cm}
\includegraphics[width=\textwidth]{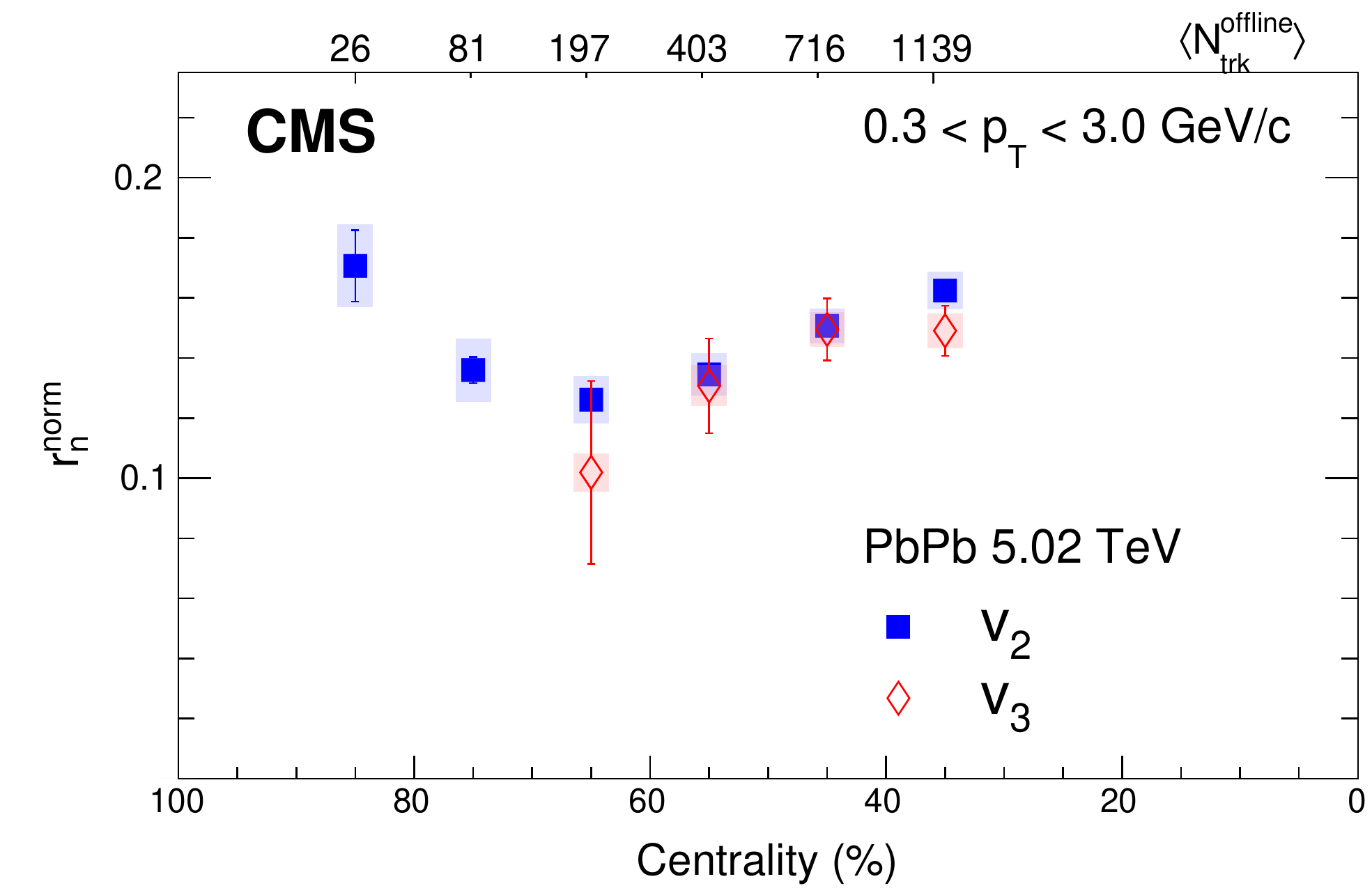}
\end{minipage}
\label{fig7}
\caption{Left: the linear slope parameters, $r^{\rm norm}$
, for $v_2$
(filled symbols) and mean $p_T$ (open symbols) as
functions of event multiplicity in p+Pb and Pb+Pb collisions at 5.02 TeV. Right: $r^{\rm norm}_2$ and $r^{\rm norm}_3$
as functions of the centrality class in
Pb+Pb collisions at 5.02 TeV. Average $N^{\rm offline}_{\rm trk}$ values for each centrality class are indicated on the top axis.
}
\label{fig7}
\end{figure}

One technical detail relevant in the $r_2$ measurement arises from the $A_{\rm ch}$ dependence of the mean $p_T$ for particles of interest. Such an effect is not expected
from the CMW signal but may occur in certain background scenario (e.g., the LCC effect). Since $v_2$ monotonically increases with $p_T$ up to $p_T$ $\sim$ 3 GeV/c, variations of the mean $p_T$ with $A_{\rm ch}$ will naturally lead to a $A_{\rm ch}$-dependent $v_2$ after integrated over the full $p_T$ range. As shown in the left panel of Fig.~\ref{fig7}, with a proper normalization, a significant fraction of the $v_2(A_{\rm ch})$ slope could be explained by the $\langle p_T \rangle(A_{\rm ch})$ slope in the CMW data of p+Pb and Pb+Pb collisions at 5.02 TeV from CMS~\cite{CMW_CMS}. As the relation between $v_2$ and $\langle p_T \rangle$ is not necessarily linear, a detailed simulation is needed to estimate
the exact fraction of the $\langle p_T \rangle$
contribution in the CMS $r_2$ results before extracting any upper limit 
on the possible residual CMW signal in the data.
On the other hand, the STAR measurements have focused on a narrow range of 
low-$p_T$ pions ($0.15 < p_T < 0.5$ GeV/$c$) to minimize the effect of 
$\langle p_T \rangle$ contributions to their results~\cite{CMW_STAR,Qi-ye3}.
It should be noted that in the LCC scenario, even for final-state particles selected with a fixed $p_T$ value, they could still originate from clusters over a wide $p_T$ range. As a function of $A_{\rm ch}$, $\langle p_T \rangle$ of clusters
will still vary, leading to a finite $r_2$ slope. In this sense, the effect of $\langle p_T \rangle$ cannot be eliminated, but is a generic LCC effect.

\subsubsection{Scaling between the second- and third-order flow harmonics}
In the LCC mechanism, a finite slope parameter for the third-order Fourier coefficient ($v_3$), $r_3$, should also arise (also true for all higher-order $v_n$) that is proportional to the baseline $v_3$ magnitude.
As both $v_2$ and $v_3$ have an approximately linear
dependence on $p_T$ at low $p_T$, a scaling
relation between the normalized $r_2$ and $r_3$ slopes is
expected,

\begin{equation}
    \langle r^{\rm norm}_2 \equiv \frac{r_2}{v^{\rm base}_2} \rangle \sim  \langle r^{\rm norm}_3 \equiv \frac{r_3}{v^{\rm base}_3} \rangle
\end{equation}
The idea is similar to the use of $\kappa_{123}$ as a data-driven background estimate for $\kappa_{112}$. The CMW is not expected to
generate a finite $r_3$ slope, as the third-order event plane has nearly no correlation with the reaction plane and/or the magnetic field direction.

The test of this prediction has been carried out by CMS, as shown in
Fig.~\ref{fig7} (right) for $r^{\rm norm}_2$
and $r^{\rm norm}_3$ as a function of centrality in Pb+Pb
collisions at 5.02 TeV~\cite{CMW_CMS}. With high precision, the normalized $r_2$ and $r_3$ slopes agree well with each other, indicating the data can be explained entirely by the LCC scenario without any CMW contribution. Future tests
in a pure-background scenario, such as the p+A data as a 
data-driven test ground and realistic model
calculations, will help further validate this scaling relation.

In the measurement of $r_2$ and $r_3$ slopes, cautions should
be taken to the possible existence of so-called nonflow 
effects (e.g., short-range correlations), which can give rise to a 
``trivial" contribution to $r_2$~\cite{trivial}. The quantitative influence of this effect to $r_n$ depends on 
details but it is generally more significant for low multiplicity events, such as low-energy, peripheral A+A and p+A collisions. The sign of 
its contribution to $r_2$ and $r_3$ is opposite. This effect can be
partially mitigated by imposing a large $\eta$ gap in measuring
$v_2$ and $v_3$ coefficients as implemented in the CMS measurement.

\subsection{Outlook}

The isobaric collisions at RHIC could benefit the search of the CMW in a similar way as that of the CME, except that the observed difference between Ru+Ru and Zr+Zr is expected to be less significant in the CMW observable than the CME one. The reason is that $r_2$ is proportional to the magnetic field, whereas $\gamma_{112}$ is proportional to the magnetic field squared,
which amplifies the difference between Ru+Ru and Zr+Zr. In spite of that, with the current statistics taken by the STAR experiment in the RHIC run 2018, there could still be a $3\sigma$ difference in $r_2$ between the two isobaric systems, if dominated by the CMW signal contribution.

The RHIC BES-II program will greatly reduce statistical uncertainties of $r_2$ measurements in Au+Au collisions at lower energies, and it is of great interest to see if $r_2$ really goes to zero at 11.5 or 7.7 GeV.
With a large $\eta$ gap, the newly-installed EPD will help suppress nonflow contribution, which could fake a sizable $r_2$  at lower energies.

The CMW consists of two chiral gapless modes traveling at the same speed~\cite{CMW}:
the right-handed (left-handed) wave transports the right-handed (left-handed) density and current in the direction
parallel (antiparallel) to the $\overrightarrow{B}$ direction.
A more general theoretical analysis~\cite{Gorbar} studied various possible collective modes based on a non-neutral-background QGP
(i.e. with nonzero $\mu$ and/or $\mu_5$) in external electric and/or magnetic fields,
and found a new type of collective motion, the chiral electric wave (CEW),
arising from CESE and propagating in parallel/antiparallel to the $\overrightarrow{E}$ field.
In symmetric collisions there should be no net electric field on average,
but asymmetric collisions like Cu+Au could provide a test ground for the CEW measurements.
\section{CVE searches in nuclear collisions}
\label{sec:CVE}
The experimental manifestation of the CVE is similar to that of the CME, except that the baryonic charge separation, instead of the electric charge separation, is induced with respect to the reaction plane.
As a result, the $\gamma$ correlators developed for the CME
search are also applicable to search for the CVE, if replacing
different electric charge combinations by combinations of baryon and anti-baryon numbers. 

A natural choice is to study correlations between protons and 
anti-protons. However, there is an ambiguity that (anti)protons
also carry electric charges so the CME may also arise, which is
indistinguishable from the CVE. Neutral baryons, such as $\Lambda$, will provide a cleaner probe to search for the baryonic charge separation effect.

Although the $\Lambda$ baryon is electrically neutral, it
carries a strange quark, which has a larger mass and thus could be 
less ``chiral''. Therefore, strange quarks may behave 
differently from up/down quarks in the chiral dynamics in 
heavy-ion collisions. As a first step to validate the behavior 
of $\Lambda$, preliminary STAR $\gamma_{112}$ measurements 
have been performed on $\Lambda$-$h^{+}$ 
($\bar{\Lambda}$-$h^{-}$) and $\Lambda$-$h^{-}$ 
($\bar{\Lambda}$-$h^{+}$) as functions of centrality in Au+Au collisions at 200 GeV~\cite{Lambda_CVE}.
In this analysis, (anti)protons have been excluded from $h^\pm$ in the correlator to avoid any possible CVE contribution.
No charge dependence is observed in this measurement, which
assures that the $\Lambda$ baryon manifests no electric charge effect 
in the $\gamma_{112}$ correlation.
Furthermore, the $\Lambda$-$h$ correlation provides a baseline for the $\Lambda$-$p$ correlation in search of the CVE. Any possible signal observed in the latter should not arise from the CME contribution.

Preliminary STAR data have been obtained on $\gamma_{112}$ of $\Lambda$-$p$ ($\bar{\Lambda}$-$\bar{p}$) and
$\Lambda$-$\bar{p}$ ($\bar{\Lambda}$-$p$) as functions of centrality in Au+Au collisions at 200 GeV~\cite{Lambda_CVE}.
The same-baryonic-charge correlation is below the
opposite-baryonic-charge correlation
from mid-central to peripheral collisions. 
This baryonic-charge separation with respect to the event plane
is consistent with the presence of a CVE signal.

Just like in the CME search, comprehensive investigations of background contributions are
necessary. In fact, similar
background effects in the CME search could also 
come into play in the CVE study.
For example, in analogy to the local charge conservation, 
the local baryonic-charge conservation could play a similar role as LCC when coupled to the elliptic flow ($v_2$).
Most of ideas and tools developed to constrain background contributions in the CME search can be easily migrated to the CVE search. The relation between $\kappa_{112}$,  $\kappa_{123}$ and $\kappa_{132}$ of $\Lambda$-$p$ correlations in both small and large systems shall shed light on the nature of background contributions. The event-shape engineering technique is applicable to eliminate the $v_2$ dependent background and determine the true CVE signal (or set an upper limit). 

As $\Lambda$ baryons are less abundantly produced than pions, higher luminosity data samples are generally required to achieve good precision for the CVE search in future programs, such as isobaric runs and BES-II at RHIC. No difference in the $\Lambda$-$p$ correlations is expected in the data of Ru+Ru and Zr+Zr collisions, since vorticity is supposed to be the same for these isobaric systems. The CVE analyses will benefit from the BES-II program, because vorticity increases at lower beam energies, and hence the CVE may yield stronger signals to be observed.
\section{Summary}
\label{sec:Outlook}
In chiral systems, the interplay of quantum anomalies with a strong magnetic field or vorticity is predicted to result in a variety of novel transport phenomena such as the chiral magnetic effect (CME), the chiral magnetic wave (CMW) and the chiral vortical effect (CVE). These phenomena probe topological properties of the QCD vacuum, and can be explored experimentally in high-energy nuclear collisions via the charge-dependent azimuthal correlations of produced hadrons from the QGP medium. In this article, we have reviewed the latest progress of experimental searches for these chiral transport effects at RHIC, BNL and the LHC, CERN over the 
past couple of decades.

The three-particle charge-dependent $\gamma$ correlator has
been introduced as the main workhorse in the search for the CME. Clear differences between the opposite- and same-sign $\gamma$ correlations are observed, which persist up to the LHC energies and down to the RHIC BES energies, with a hint of diminishing at 7.7 GeV. This is in line with expectation of the CME picture. 
However, possible background contributions such as resonance decays, local charge conservation and transverse momentum conservation are identified, which could explain a sizable fraction of, if not all, the observed charge separation. Their contributions have to be carefully quantified before any claim of the discovery of the CME can be made. Intensive efforts have been made to constrain the CME backgrounds with phenomenological models and data-driven approaches.

We reviewed several data-driven background studies including
the small-system (p+A, d+A) data, the $\gamma$ correlator with respect to higher-order event planes, where no CME signals are expected; and the event-shape engineering technique, with the goal of varying
the signal and background contributions in a controlled way.
Present results suggest that little, if any, contribution of the CME signal is present in the experimental observable of Pb+Pb collisions at LHC energies, with an upper limit less than only a few \%. However, future work of applying these studies to lower RHIC energies at BES-II and isobaric programs is still promising to identify a possible CME signal that is expected to be larger in that energy regime.

The status of experimental searches for the CMW is reviewed, with the main focus on the measurement of charge-dependent elliptic flow as a function of the observed event charge asymmetry. Positive signals have been observed at both RHIC and LHC energies, supporting the CMW picture. Similar to the CME search,
efforts have been made to understand possible contributions of background sources using small system
data and higher-order flow harmonics, where no CMW signals are expected. Striking similarity between the p+Pb and Pb+Pb data at LHC energies suggest the dominance of backgrounds, while p+Au and d+Au results at RHIC energies seem to be consistent with zero, indicating a possible CMW signal in Au+Au collisions.
Future programs of isobaric collisions and BES-II at RHIC will help clarify if a CMW signal indeed exists and its potential beam energy dependence.

Experimental searches for the CVE signal has been performed with the same $\gamma$ correlator but replacing charged particles by (anti)$\Lambda$ and (anti)proton in Au+Au collisions at 200 GeV. A difference between the opposite- and same-baryonic-charge correlations is observed 
from peripheral to mid-central collisions, consistent with the CVE expectation.
However, just like in the CME search, extensive work to quantitatively understand background contributions is still needed to draw a definitive conclusion on the observation of the CVE signal.

\section*{DISCLOSURE STATEMENT}
The authors are not aware of any affiliations, memberships, funding, or financial holdings that
might be perceived as affecting the objectivity of this review. 

\section*{ACKNOWLEDGMENTS}
Gang Wang is supported by the U.S. Department of Energy, Office of Nuclear Physics, under the Grant DE-FG02-88ER40424. Wei Li is supported by the U.S. Department of Energy, Office of Nuclear Physics, under the Grant DE-SC0005131 and the Welch Foundation (Grant No. C-1845).


\end{document}